\documentclass[aps,prl,reprint,superscriptaddress,longbibliography]{revtex4-2}
\usepackage[T1]{fontenc}
\usepackage{amsmath,amssymb,bm,graphicx,booktabs,array,mathrsfs}
\usepackage[colorlinks=true,linkcolor=blue,citecolor=blue,urlcolor=blue]{hyperref}
\newcommand{\BZ}{\mathrm{BZ}}
\newcommand{\kb}{k_{\mathrm B}}
\newcommand{\BA}{B_{\mathrm A}}
\newcommand{\av}[1]{\left\langle #1\right\rangle_{\BZ}}
\newcommand{\ii}{\mathrm i}
\begin{document}
\title{Intrinsic Orbital Nernst Effect in Real Photonic Bands}
\author{Tao Zhu}
\email{zhutao@tiangong.edu.cn}
\affiliation{School of Electronic and Information Engineering, Tiangong University, Tianjin, China}
\date{11 October 2026}
\begin{abstract}
We predict an intrinsic orbital Nernst effect in reciprocal photonic lattices with real Bloch modes. In the collisionless regime, a thermal gradient generates a transverse angular-displacement response although band angular moments and ordinary Berry curvatures vanish. The response is carried by angular-density dipoles formed through virtual mixing of distinct spatial modes. Their correlation with band velocity survives thermal averaging and is independently recovered from the angular-density readout of unitarily evolved wave packets. Two controls expose the mechanism: a fixed-generator two-level system has zero antisymmetric intrinsic response, and flattening the dispersion suppresses transport without changing the eigenstates. A canonical scalar Maxwell calculation realizes the effect through symmetry-selected virtual channels, even when one band dominates the thermal population. These findings establish an orbital-dipole mechanism for intrinsic thermal angular transport in real photonic bands.
\end{abstract}
\maketitle

Can a temperature gradient drive transverse orbital angular transport in photonic bands whose eigenstates carry neither angular momentum nor Berry curvature? Optical Hall effects usually reveal polarization-dependent deflection or the displacement of vortex beams \cite{Onoda2004,Bliokh2015,Athira2021}. Real photonic bands pose a different problem: the familiar band moment and anomalous-velocity ingredients both vanish. Their spatial mode structure nevertheless contains local angular information that can become visible in thermal transport.

Orbital texture sustains electronic Hall and Nernst responses despite orbital quenching \cite{Bernevig2005,Go2018,Salemi2022}, with recent experiments probing electronic orbital Nernst transport \cite{Li2026,Masuda2026}. Thermal angular currents also occur in magnons \cite{Cheng2016,Zyuzin2016,Weissenhofer2026} and phonons \cite{Park2020,Lopez2026}. In photonics, thermal Hall transport \cite{Ben2016}, transverse-spin emission from reciprocal temperature gradients through nonequilibrium mode mixing \cite{Chen2022}, and nonreciprocal spin--heat conversion \cite{Ben2025,Ben2026} have opened related routes. The near-field photon Nernst effect concerns a photon-induced electronic response \cite{Dehaghi2026}. We address the intrinsic transport of spatial-mode orbital angular momentum \cite{Milicevic2017} within a reciprocal lattice, distinct from the spin of emitted radiation.

\begin{figure*}[t]
 \includegraphics[width=\textwidth]{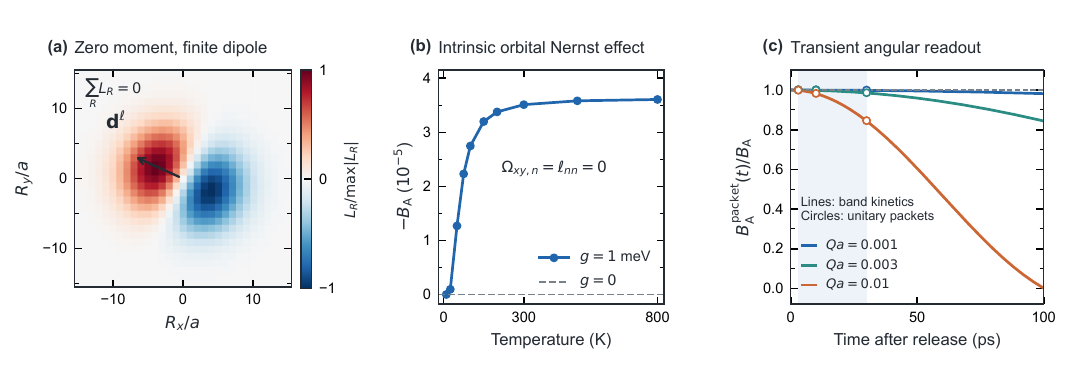}
 \caption{Intrinsic orbital Nernst effect carried by angular-density dipoles. The local model has $g=1$ meV and $\delta=2.5$ meV, except for the $g=0$ control. (a) Cell angular momentum $L_R$ of a normalized lowest-band Gaussian packet: opposite lobes give zero total moment but a finite first moment; the arrow indicates its direction. (b) Thermal response despite $\Omega_{xy,n}=\ell_{nn}=0$; the decoupled limit has zero antisymmetric response. (c) Transient coefficient from orthogonal gratings at $300$ K, normalized to the bulk value. Lines: band kinetics. Open circles: direct unitary evolution and resolved angular-density readout, with momentum width $\sigma_q=0.04$. Here $Qa$ is the dimensionless grating wave number. The shaded $3$--$30$ ps interval lies between the interband and grating-transit times for $Qa=10^{-3}$.}
 \label{fig:effect}
\end{figure*}

Here we predict such an intrinsic orbital Nernst effect and identify its microscopic carrier: an angular-density dipole. A wave packet can have zero total angular momentum yet contain opposite angular-density lobes with a finite first moment [Fig.~\ref{fig:effect}(a)]. Virtual mode mixing builds this dipole, and its correlation with band velocity produces a transverse thermal response. We demonstrate the effect in a reciprocal three-mode model, resolve its microscopic polarization dynamics, and realize the same geometric mechanism in a canonical scalar Maxwell field. Two decisive controls---removing an independent virtual channel and flattening the dispersion at fixed eigenstates---reveal when the effect switches off.

\emph{Angular dipoles and thermal transport.}---The Bloch Hamiltonian obeys $h(\bm q)|u_n\rangle=E_n|u_n\rangle$, with positive energies $E_n$ and normalized band modes $|u_n\rangle$. Here $q_i=a_i k_i$ is dimensionless Bloch momentum, $a_i$ is the lattice period, and $\partial_i\equiv\partial/\partial q_i$, for $i=x,y$; the local model has $a_x=a_y=a$. A constant angular generator $\ell=L_z/\hbar$ acts in a fixed spatial-orbital basis. We use the time derivative of the angular displacement $\{R_i,L_z\}/2$, where $R_i$ is cell position and braces denote an anticommutator, including its coherent torque contribution \cite{Shi2006,Xiao2021,Koyama2025}. For a Bose distribution with zero chemical potential, define
\begin{equation}
 J_i^L=-\alpha^L_{ij}\nabla_jT,\qquad
 \BA=\frac{\alpha^L_{yx}-\alpha^L_{xy}}{2\kb}.
 \label{eq:definition}
\end{equation}
Here $J_i^L$ is angular-current density, $\nabla_jT$ the real-space temperature gradient, and $\kb$ Boltzmann's constant. In two dimensions $J_i^L$ has units of energy per length, so $\alpha^L_{ij}/\kb$ and $\BA$ are dimensionless. The latter isolates the orientation-independent transverse response; the symmetric coefficient $B_{\rm S}=(\alpha^L_{yx}+\alpha^L_{xy})/(2\kb)$ is removed by averaging orthogonal drive directions.

For an isolated band, let $|D_iu_n\rangle=(1-|u_n\rangle\langle u_n|)|\partial_iu_n\rangle$. Its orbital-density dipole \cite{Culcer2004} is
\begin{equation}
 d^\ell_{n,i}=\operatorname{Im}\langle D_iu_n|\ell|u_n\rangle.
 \label{eq:dipole}
\end{equation}
The physical internal angular dipole is $\hbar a_i d^\ell_{n,i}$. A real Hamiltonian and imaginary antisymmetric $\ell$ enforce the band moment $\ell_{nn}=\langle u_n|\ell|u_n\rangle=0$ and ordinary Berry curvature $\Omega_{xy,n}=0$, while allowing $\bm d^\ell_n\ne0$. Including torque multipoles and subtracting equilibrium circulation \cite{Luttinger1964,Qin2011,Matsumoto2011}, the thermal response reduces to
\begin{equation}
 \BA=\frac12\av{\sum_n w_n
 [ (\partial_xE_n)d^\ell_{n,y}-(\partial_yE_n)d^\ell_{n,x}]},
 \label{eq:velocitydipole}
\end{equation}
where $n_n=(e^{E_n/(\kb T)}-1)^{-1}$ is the Bose occupation, $w_n=E_nn_n(1+n_n)/(\kb T)^2$ the thermal weight, and $\av{f}=\int_\BZ d^2q\,f/(2\pi)^2$ the first-Brillouin-zone average. Since $v_{n,i}=a_i\partial_iE_n/\hbar$, Eq.~(\ref{eq:velocitydipole}) expresses a thermally selected transverse correlation between band velocity and angular dipole. It needs no diagonal band angular moment.

The coefficient uses the collisionless order of limits: long wavelength first, then relaxation removed before the drive frequency $\omega_{\rm d}$ tends to zero, with $\gamma\ll\hbar|\omega_{\rm d}|\ll\Delta_{\rm dir}$. Here $\gamma$ is relaxation energy and $\Delta_{\rm dir}$ the minimum separation of bands at the same momentum. Prepare a weak thermal grating $T(x,0)=T_0+\delta T\cos(Qx)$, where $Q$ is a physical wave number. After release, the packets transport angular polarization, the first moment of angular density per area; its transverse rate approaches $-\alpha^L_{yx}\nabla_xT$ before grating transit. Antisymmetrizing orthogonal grating responses defines $B_{\rm A}^{\rm packet}(t)$, which approaches $\BA$ [Fig.~\ref{fig:effect}(c)]. Direct unitary lattice evolution independently recovers this response from packet centers and angular-density first moments, without a band-dipole input. The readout resolves individual packets; an unresolved total-density moment is a different observable. Its derivation and the distinct stationary relaxation limit are given in the Supplemental Material \cite{SM}.

\begin{figure*}[t]
 \includegraphics[width=\textwidth]{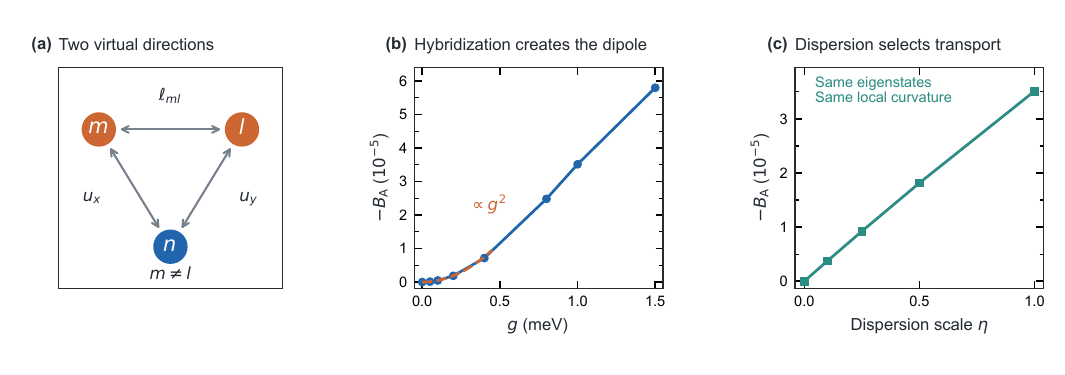}
 \caption{Controls of the orbital-dipole mechanism. (a) Distinct virtual modes $m\ne l$ connect the velocity and angular vertices of band $n$; double-headed links denote Hermitian matrix elements. (b) Hybridization activates the antisymmetric response, with a quadratic onset (dashed guide). (c) Flattening $E_{n,\eta}=\bar E_n+\eta(E_n-\bar E_n)$ at fixed eigenstates suppresses the response while preserving its local geometric kernel. Here $\bar E_n=\av{E_n}$; $\eta=1$ gives the original dispersion and $\eta=0$ flat bands. Both response panels use the local model at $300$ K and $\delta=2.5$ meV; panel (c) has $g=1$ meV.}
 \label{fig:mechanism}
\end{figure*}

\emph{Virtual mixing creates the dipole.}---Integration by parts gives an equivalent expression that exposes the participating modes:
\begin{align}
 \BA&=\av{\sum_n c_1(n_n)\mathcal F_n},\nonumber\\
 \mathcal F_n&=-\operatorname{Im}\langle D_xu_n|\ell|D_yu_n\rangle \nonumber\\
 &=-\operatorname{Im}\sum_{m,l\ne n}
 \frac{(u_x)_{nm}\ell_{ml}(u_y)_{ln}}
 {(E_n-E_m)(E_n-E_l)},
 \label{eq:kernel}
\end{align}
where $u_i=\partial_ih$ is the velocity vertex in energy units, with matrix elements in the band basis. The function $c_1(n)=(1+n)\ln(1+n)-n\ln n$ is the entropy of a bosonic mode in units of $\kb$. The kernel is half the curl of $\bm d^\ell_n$. Two independent virtual directions, joined by the angular vertex, create this oriented mode area [Fig.~\ref{fig:mechanism}(a)]. For the real-band quantum metric $\mathsf g_{ij,n}=\langle D_iu_n|D_ju_n\rangle$ \cite{Provost1980}, $|\mathcal F_n|\le\|\ell\|\sqrt{\det\mathsf g_n}$, where $\|\ell\|$ is the operator norm.

A real two-component eigenvector has only one tangent direction: its two derivatives are parallel, so the antisymmetric kernel vanishes. The cancellation is stronger than this real-band picture. For any smooth isolated two-level Hamiltonian and constant traceless local generator, the Berry-moment term cancels the dipole curl even for complex eigenstates. Thus the \emph{antisymmetric intrinsic response} is zero; symmetric tensor components may remain finite. A third spatial mode can unlock a second tangent direction without itself being appreciably occupied. Eliminating that mode also transforms the position and angular operators, so the fixed-generator theorem does not apply unchanged to the downfolded theory (Supplemental Material).

A reciprocal lattice of an even $s$ mode and two odd $p_x,p_y$ modes realizes the mechanism. Using the phase choice $(is,p_x,p_y)$,
\begin{equation}
 \begin{gathered}
 h=\begin{pmatrix}
 \epsilon_s&g\sin q_x&g\sin q_y\\
 g\sin q_x&\epsilon_p+d_z+\delta&d_x\\
 g\sin q_y&d_x&\epsilon_p-d_z-\delta
 \end{pmatrix},\quad \ell=0\oplus\sigma_y,
 \end{gathered}
 \label{eq:model}
\end{equation}
where $\epsilon_s=E_s+b_s(\cos q_x+\cos q_y)$, $\epsilon_p=E_p+b_p(\cos q_x+\cos q_y)$, $d_z=A(\cos q_x-\cos q_y)$, and $d_x=C\sin q_x\sin q_y$. Here $E_s,E_p$ set the $s$ and mean $p$ onsite energies; $b_s,b_p,A$ set dispersion, $C$ mixes the two $p$ modes, and $g$ couples $s$ to $p$. The Pauli matrix $\sigma_y$ acts on the $p$ pair. Combined inversion and time reversal permit the real frame. Time reversal is $\operatorname{diag}(-1,1,1)\mathcal K$, with $\mathcal K$ complex conjugation; angular current is even because velocity and angular momentum are both odd. The reciprocal onsite splitting is $2\delta$.

With a circular-cavity calibration setting the $p$ scales, we use $E_s=14$, $b_s=0.3$, $E_p\simeq19.936$, $b_p=0.71898$, $A=0.98492$, and $C=0.18712$ meV (full constants in the Supplemental Material). At $g=1$ meV and $\delta=2.5$ meV the bands are separated by a minimum direct gap of $1.06$ meV, and
\begin{equation}
 \BA(300\ \mathrm K)\simeq-3.51\times10^{-5}.
 \label{eq:result}
\end{equation}
Figure~\ref{fig:effect}(b) shows the temperature dependence. At $300$ K, turning off $g$ restores $\BA=0$, although $B_{\rm S}\simeq-2.57\times10^{-5}$ remains finite. Weak mixing gives $\BA\simeq-4.7\times10^{-5}(g/\mathrm{meV})^2$ [Fig.~\ref{fig:mechanism}(b)]. The sign of $g$ is removable by a basis phase.

For a real three-component mode $\bm u_n$, let $\bm e_s=(1,0,0)$ select its $s$ component $u_{s,n}$. Then $\mathcal F_n=\bm e_s\cdot(\partial_x\bm u_n\times\partial_y\bm u_n)=u_{s,n}\mathcal E_n$, where $\mathcal E_n=\bm u_n\cdot(\partial_x\bm u_n\times\partial_y\bm u_n)$ is locally the Euler curvature of the plane orthogonal to $\bm u_n$ in mode space \cite{Ahn2019,Bouhon2020}. Near an $s$ mode, its two $O(g)$ admixture amplitudes $\beta_x,\beta_y$ give $\mathcal F_s=\partial_x\beta_x\partial_y\beta_y-\partial_y\beta_x\partial_x\beta_y+O(g^4)$, explaining the quadratic onset. The response is geometric and need not be topological: our finite-response model has a trivial complementary plane (Supplemental Material).

\emph{Dispersion selects the thermal response.}---The second control follows directly from Eq.~(\ref{eq:velocitydipole}). Flattening each band while keeping its eigenstates leaves the dipole and its curl intact, but cancels their integrated thermal response [Fig.~\ref{fig:mechanism}(c)]. The residual falls below $2\times10^{-19}$ at $300$ K. Virtual mixing creates the angular texture; dispersion prevents cancellation between regions of opposite dipole curl.

\begin{figure*}[t]
 \includegraphics[width=\textwidth]{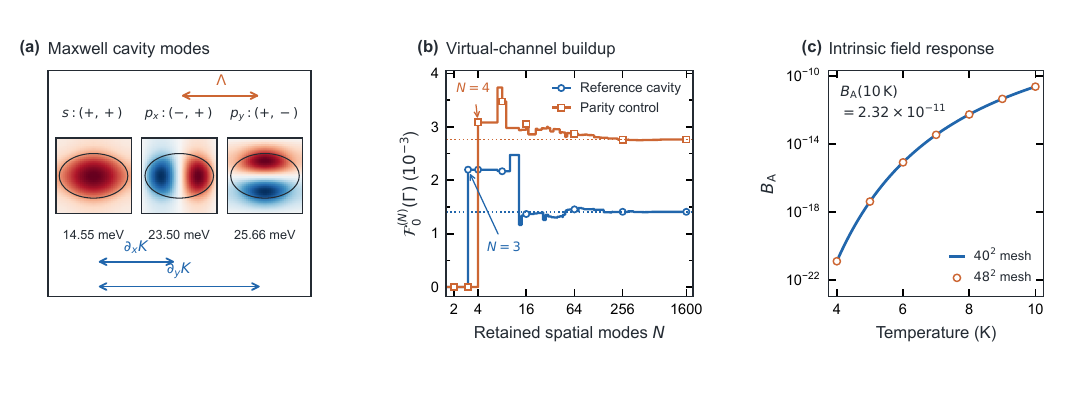}
 \caption{Realization in a canonical scalar Maxwell field. (a) Calculated modes at $\Gamma$ ($\bm q=0$). Ellipses mark the cavity boundary; parities refer to the $x,y$ mirrors. Red/blue denote positive/negative real amplitudes, normalized separately. Links denote matrix elements of $\partial_iK$ and $\Lambda$. (b) Lowest-band kernel $\mathcal F_0^{(N)}(\Gamma)$, with its virtual sum restricted to the lowest $N$ spatial modes, including the ground mode. The reference activates at $N=3$; the parity control requires $N=4$. Dotted lines mark all 1600 modes of the $40\times40$ mesh. (c) Unscaled response from 32 thermally weighted bands; symbols show a $48\times48$ spatial mesh. At $10$ K, $\BA\simeq2.32\times10^{-11}$ and the lowest band supplies $99.7\%$ of this response. Geometries and mode-pair contributions are detailed in the Supplemental Material.}
 \label{fig:field}
\end{figure*}

\emph{The effect in a Maxwell field.}---We solve the lossless transverse-magnetic (TM, electric field along $z$) problem for vacuum cavities of semiaxes $R_x,R_y$ in an ideal Drude host. With spatial coordinates in units of $R_x$, $K=-\nabla_{\rho}^2+\kappa^2\chi_{\rm host}$, where $\chi_{\rm host}$ is one in the host and zero in vacuum, and $\kappa=\omega_pR_x/c$ uses the host plasma frequency $\omega_p$ and vacuum light speed $c$. Define the spatial rotation derivative $D_\varphi=\rho_x\nabla_{\rho,y}-\rho_y\nabla_{\rho,x}$ about the cavity center. The angular observable integrates the Hermitian bilinear associated with $-\Pi D_\varphi\Phi$ \cite{Bliokh2010,Bliokh2017,Go2024}; $\Phi$ is the scalar vector-potential coordinate and $\Pi$ its conjugate momentum, with canonical normalization given in the Supplemental Material. Weighting the cavity angular momentum by cell position defines its displacement. Field correlations provide an in-principle readout, building on spatial-mode tomography \cite{Guillot2025}.

On the spatial mesh, let $G=-C_{\rm cav}D_\varphi$, with $C_{\rm cav}$ the diagonal vacuum window, and decompose it into $\Lambda=i(G-G^T)/2$ and $S=(G+G^T)/2$. Normalized scalar modes obey $K\psi_n=\Omega_n^2\psi_n$ with Bloch boundary conditions, where $\Omega_n=\omega_nR_x/c$ is dimensionless, $E_n=E_0\Omega_n$, and $E_0=\hbar c/R_x$. A local canonical reduction \cite{Koyama2025,Colpa1978} includes the non-diagonal angular vertex, torque multipoles, and circulation subtraction. Summing both signs of the virtual canonical frequencies yields
\begin{equation}
 d^{\rm can}_{n,i}=\operatorname{Im}\langle D_i\psi_n|\Lambda|\psi_n\rangle
              +\frac{S_{nn}}{2\Omega_n}\partial_i\Omega_n.
 \label{eq:canonicaldipole}
\end{equation}
Here $D_i$ is the same band-orthogonal momentum derivative as above, and $S_{nn}=\langle\psi_n|S|\psi_n\rangle$. The second term is parallel to band velocity and contributes zero integrated antisymmetric response. The remaining dipole has the structure of Eq.~(\ref{eq:dipole}), with $\mathcal F_n=-\operatorname{Im}\langle D_x\psi_n|\Lambda|D_y\psi_n\rangle$. Positive-frequency thermal modes are counted once; the complete spatial space enters virtually.

For $R_y/R_x=0.65$, $a_x=a_y=2.2R_x$, and $\kappa=4$, the lowest band has a minimum sampled direct gap of $4.67$ meV and a unique energy minimum at the Brillouin-zone center $\Gamma$. The three lowest modes have parities $(+,+),(-,+),(+,-)$ [Fig.~\ref{fig:field}(a)]. The $x$ and $y$ velocity vertices access the two odd sectors, and $\Lambda$ connects them. Their cumulative contribution activates at three retained modes [Fig.~\ref{fig:field}(b)]. Higher modes reduce the kernel from $\mathcal F_0^{(3)}(\Gamma)\simeq2.20\times10^{-3}$ to $1.41\times10^{-3}$, a $35.9\%$ reduction. In a parity control with $\kappa=6$, $a_x=2.2R_x$, and $a_y=1.5R_x$, the third mode is even. The three-mode kernel vanishes, and adding the fourth, $(+,-)$ mode activates it. Thus symmetry of the virtual modes, beyond their number, controls the effect.

We set $E_0=\hbar c/R_x\simeq6.682$ meV by assigning $20$ meV to the isolated circular-cavity $p$ mode at $\kappa=4$ (Supplemental Material). The direct field integral gives $\BA\simeq2.32\times10^{-11}$ at $10$ K, changing by $0.13\%$ between $40\times40$ and $48\times48$ spatial meshes [Fig.~\ref{fig:field}(c)]. The lowest band supplies $99.7\%$ of this response. For its energy $E(\bm q)=E_{\min}+\bm q^TM\bm q/2+O(q^4)$, $M$ is the positive Hessian at $\Gamma$, in energy units. Its low-temperature response obeys
\begin{equation}
 \BA(T)\sim C_0e^{-E_{\min}/(\kb T)},\qquad
 C_0=\frac{E_{\min}\mathcal F_0(\Gamma)}{2\pi\sqrt{\det M}}.
 \label{eq:activation}
\end{equation}
Here $\mathcal F_0$ labels the lowest-band kernel and $C_0$ is dimensionless. We obtain $E_{\min}\simeq14.55$ meV and $C_0\simeq0.0131>0$, verified by an independent asymptotic integral. This prefactor establishes the effect as $T\to0$; the unscaled curve in Fig.~\ref{fig:field}(c) reports its actual magnitude and is evaluated independently. The finite response is small because of thermal activation. One occupied band can therefore carry the effect while borrowing its dipole from several virtual modes. Disconnecting the cells restores zero response.

We have predicted an intrinsic orbital Nernst effect in real photonic bands and traced it to angular-density dipoles generated by virtual spatial-mode mixing. The velocity--dipole correlation links its geometric kernel to thermal transport and to an explicit transient polarization response. Two-level cancellation and fixed-eigenstate flattening expose the two ingredients, while the scalar Maxwell realization shows how cavity symmetry opens the required virtual channels. This mechanism establishes a route to reciprocal photonic orbitronics without a band angular moment or ordinary Berry curvature. Optimizing the mode energies and coupling the angular observable to reservoirs and probes are natural next steps; photon nonequilibrium Green's functions provide the corresponding energy, momentum, and angular vertices \cite{Zhang2022,Wang2023}.

\begin{acknowledgments}
\interlinepenalty=10000
This work is supported by the National Key R$\&$D program of China (Project No. 2022YFA1204000) and the National Natural Science Foundation of China (Grant No. 12204346).
\end{acknowledgments}

% Supplemental Material: same source file, independent S numbering.
\clearpage
\onecolumngrid
\setcounter{secnumdepth}{2}
\setcounter{section}{0}
\setcounter{subsection}{0}
\setcounter{equation}{0}
\setcounter{figure}{0}
\setcounter{table}{0}
\renewcommand{\thesection}{S\arabic{section}}
\renewcommand{\thesubsection}{\thesection.\arabic{subsection}}
\makeatletter
\def\p@subsection{}
\def\p@subsubsection{}
\makeatother
\numberwithin{equation}{section}
\renewcommand{\thefigure}{S\arabic{figure}}
\renewcommand{\thetable}{S\arabic{table}}
\renewcommand{\theHsection}{supp.\arabic{section}}
\renewcommand{\theHsubsection}{supp.\arabic{section}.\arabic{subsection}}
\renewcommand{\theHequation}{supp.\arabic{section}.\arabic{equation}}
\renewcommand{\theHfigure}{supp.\arabic{figure}}
\renewcommand{\theHtable}{supp.\arabic{table}}
\normalsize
\setlength{\parskip}{3pt}
\pdfbookmark[0]{Supplemental Material}{supplement-start}

\begin{center}
{\large\bfseries Supplemental Material for\\[3pt]
Intrinsic Orbital Nernst Effect in Real Photonic Bands}\\[8pt]
Tao Zhu\\[3pt]
{\small School of Electronic and Information Engineering, Tiangong University, Tianjin, China}\\[3pt]
\href{mailto:zhutao@tiangong.edu.cn}{\texttt{zhutao@tiangong.edu.cn}}\\[6pt]
11 October 2026
\end{center}

The main text develops the effect through three steps: the angular-density dipole and its transient response (Fig.~1), the virtual-mixing and dispersion controls (Fig.~2), and the canonical scalar Maxwell realization (Fig.~3). This supplement supplies their derivations and numerical details. Sections~\ref{supp:sec:current}--\ref{supp:sec:symmetry} derive the local-mode response and selection rules. Sections~\ref{supp:sec:canonical}--\ref{supp:sec:activation} establish the field-current formula, its low-temperature response, and the symmetry-resolved virtual-mode sum. Section~\ref{supp:sec:calibration} documents the mode-model calibration. Sections~\ref{supp:sec:position}--\ref{supp:sec:covariant} explain how projection changes the operators. Section~\ref{supp:sec:thermalscope} derives the thermal-grating dynamics and field-correlation readout.

\section{Observable, current, and thermal convention}
\label{supp:sec:current}
The effective lattice Hamiltonian is $H=\sum_{\bm q}a^\dagger_{\bm q}h(\bm q)a_{\bm q}$ with three positive-frequency modes per cell. Its angular observable is $L_z=\hbar\ell$, where $\ell=0\oplus\sigma_y$ in the anchored $(is,p_x,p_y)$ basis. The eigenmodes $p_x\pm\ii p_y$ have angular-generator eigenvalues $\pm1$ (with their sign fixed by the convention for $\ell$). This local mode observable commutes with the lattice cell position $R_i$. It is not assumed to commute with the Hamiltonian.

Let $\tau^L=(\ii/\hbar)[H,L_z]$. The time derivative of the symmetrized orbital displacement illustrates the distinction from the conventional current:
\begin{equation}
 \frac{d}{dt}\frac{\{R_i,L_z\}}2
 =\frac{\{v_i,L_z\}}2+\frac{\{R_i,\tau^L\}}2.
 \label{supp:eq:proper}
\end{equation}
The bulk band formulation of this current requires torque multipoles, including the torque quadrupole, and subtraction of the equilibrium circulating contribution \cite{Shi2006,Xiao2021,Koyama2025}. We calculate its clean intrinsic term. Equation~(\ref{supp:eq:proper}) does not assert that crystal angular momentum is locally conserved at every boundary or that the result equals the torque on a detector.

Our force convention is $J_i^L=-\alpha^L_{ij}\nabla_jT$, with $\nabla_jT$ the real-space temperature gradient. For $g_n=\kb T\ln(1-e^{-E_n/(\kb T)})$,
\begin{equation}
 -\frac{\partial g_n}{\partial T}=\kb c_1(n_n),\qquad
 c_1(n_n)=\frac{z_n}{e^{z_n}-1}-\ln(1-e^{-z_n}),\quad z_n=\frac{E_n}{\kb T}.
\end{equation}
Only positive-frequency modes are counted, and the thermal photon chemical potential is zero. We use $\kb=0.08617333262145$ meV/K. With dimensionless $q_i=a_i k_i$, lattice periods $a_i$, and $u_i=\partial_{q_i}h$, the physical velocity vertex is $v_i=a_i u_i/\hbar$; the local model has $a_x=a_y=a$. Unadorned $\partial_i$ denotes a momentum derivative, whereas $\nabla_{\rho,i}$ below differentiates the spatial coordinate $\rho_i$. The angular-current factor of $\hbar$, two momentum derivatives, and the two-dimensional momentum measure give the dimensionless normalization $\alpha^L_{ij}/\kb$ used in the main text. The physical angular current density has units of energy per length, equivalently angular momentum per time per length.

For a constant local $\ell$, the adopted intrinsic formula is
\begin{equation}
 \frac{\alpha^L_{ij}}{\kb}=\av{\sum_n c_1(n_n)
 [\Omega_{ji,n}\ell_{nn}+\partial_jd^\ell_{n,i}]},
 \qquad
 d^\ell_{n,i}=\operatorname{Im}\sum_{m\ne n}
 \frac{(u_i)_{nm}\ell_{mn}}{E_n-E_m}.
 \label{supp:eq:general}
\end{equation}
The position-commuting local-operator assumption is essential. Replacing $\ell$ in this expression by an arbitrary momentum-dependent projected continuum operator without rederiving the current is not justified. The present model uses a number-conserving bosonic Hamiltonian; no bosonic Bogoliubov doubling or extra factor of two is present.

The use of Bose rather than Fermi statistics can be made explicit. For one-body bilinears, the operator identity
\begin{equation}
 [a_n^\dagger a_m,a_l^\dagger a_p]
 =\delta_{ml}a_n^\dagger a_p-\delta_{np}a_l^\dagger a_m
\end{equation}
has the same form for both statistics. Consequently the retarded correlator has the same matrix structure,
\begin{equation}
 \chi^R_{AB}(\omega)=\sum_{nm}
 \frac{(n_n-n_m)A_{nm}B_{mn}}{\hbar\omega+E_n-E_m+\ii0^+}.
\end{equation}
For the current definition in Eq.~(\ref{supp:eq:proper}), the established torque-multipole identity combines the conventional-current and torque kernels into $\Omega_{ji,n}\ell_{nn}+\partial_jd^\ell_{n,i}$ \cite{Xiao2021}. Changing statistics replaces the occupation and grand potential, without changing that one-body geometry. Equilibrium-current subtraction, essential in bosonic thermal transport \cite{Qin2011,Matsumoto2011}, then gives $-\partial_Tg_n=\kb c_1(n_n)$ and Eq.~(\ref{supp:eq:general}). This use of a known transport framework is an input to the present model analysis, not a claim of a new general current theory. A bosonic treatment with a doubled Bogoliubov space is discussed in Ref.~\cite{Koyama2025}; its redundant sectors are absent here.

The local continuity condition can also be checked directly. Set $\bm q_\pm=\bm q\pm\bm Q/2$ and $\rho_\ell(\bm Q)=\sum_{\bm q}a^\dagger_{\bm q_-}\ell a_{\bm q_+}$. Then
\begin{align}
 \dot\rho_\ell(\bm Q)&=\frac{\ii}{\hbar}\sum_{\bm q}a^\dagger_{\bm q_-}
 (h_-\ell-\ell h_+)a_{\bm q_+},\\
 h_-\ell-\ell h_+&=[h,\ell]-Q_i\frac{\{u_i,\ell\}}2
 +\frac{Q_iQ_j}{8}[\partial_i\partial_jh,\ell]+O(Q^3).
\end{align}
This separates the conventional current from coherent torque and shows why second-order torque information can enter a first-gradient current. With $\vartheta=\ii[h,\ell]$, the band torque-source curvature satisfies
\begin{equation}
 \mathcal C^{\vartheta}_{n,j}
 =-2\operatorname{Im}\sum_{m\ne n}
 \frac{(u_j)_{nm}\vartheta_{mn}}{(E_n-E_m)^2}
 =\partial_j\ell_{nn}=0.
 \label{supp:eq:torquesource}
\end{equation}
The last equality holds pointwise for our real $h$ and imaginary antisymmetric $\ell$. Thus the bulk torque monopole does not obstruct the adopted transport construction, although torque dipoles and quadrupoles still contribute. This argument does not require the matrix equality $h(\bm q)=h(-\bm q)$; our inversion acts through an orbital parity matrix. The sign convention in Eq.~(\ref{supp:eq:general}) follows Eq.~(11) of Ref.~\cite{Xiao2021} with $J=-\alpha\nabla T$. A Luttinger source \cite{Luttinger1964}, $\delta h=\{\psi,h\}/2$, represents the transport drive through $\nabla\psi=\nabla T/T$ after equilibrium subtraction. The opposite sign describes equilibrium compensation, not this transport drive.

\section{Dipole curl and the two-band cancellation}
\label{supp:sec:selection}
For real $h$ the eigenvectors can be chosen locally real away from degeneracies. With $\ell$ imaginary antisymmetric, $\ell_{nn}=0$, $\Omega_{ij,n}=0$, and $\langle u_n|\partial_i u_n\rangle=0$. Differentiating the dipole gives
\begin{align}
 \partial_xd^\ell_{n,y}-\partial_yd^\ell_{n,x}
 &=\operatorname{Im}\bigl[\langle\partial_yu_n|\ell|\partial_xu_n\rangle
             -\langle\partial_xu_n|\ell|\partial_yu_n\rangle\bigr]\\
 &=-2\operatorname{Im}\langle D_xu_n|\ell|D_yu_n\rangle.
 \label{supp:eq:curl}
\end{align}
The mixed second derivatives cancel. The covariant derivatives can be evaluated without differentiating eigenvector signs:
\begin{equation}
 |D_i u_n\rangle=\sum_{m\ne n}|u_m\rangle
 \frac{(u_i)_{mn}}{E_n-E_m}.
\end{equation}
Equations~(\ref{supp:eq:general}) and (\ref{supp:eq:curl}) then give the interband kernel in the main text. Its real-model matrix elements with $m=l$ vanish, so a two-band subspace cannot contribute a smooth kernel. Turning off the $s$--$p$ coupling in a three-band model leaves exactly such a block, plus a decoupled scalar band.

A useful stronger algebraic check is a general isolated two-level Hamiltonian $h=\epsilon+\bm d\cdot\bm\sigma$ with a constant traceless generator $\ell=\bm l\cdot\bm\sigma$. Set $\bm n=\bm d/|\bm d|$ and label the bands by $s=\pm1$. Direct projector algebra yields
\begin{align}
 \Omega_{xy,s}&=-\frac{s}{2}\bm n\cdot(\partial_x\bm n\times\partial_y\bm n),
 &\ell_{ss}&=s\bm l\cdot\bm n,\\
 d^\ell_{s,i}&=\frac12\bm l\cdot(\bm n\times\partial_i\bm n).
\end{align}
Since derivatives of a unit vector are tangent to its sphere,
\begin{equation}
 \frac12(\partial_xd^\ell_{s,y}-\partial_yd^\ell_{s,x})
 =\frac12(\bm l\cdot\bm n)\bm n\cdot(\partial_x\bm n\times\partial_y\bm n)
 =-\Omega_{xy,s}\ell_{ss}.
\end{equation}
Thus the antisymmetric kernel in Eq.~(\ref{supp:eq:general}) cancels for this isolated two-level problem as well. An identity component of the observable would retain an ordinary Berry-curvature response. Conserved-spin-current cancellations have precedents in Rashba transport \cite{Xiao2021}; the identity here is pointwise in the clean geometric kernel of a smooth isolated two-level problem. The weak-disorder stationary cancellation discussed in that reference additionally involves a collision contribution and a different order of limits.

At a real two-band degeneracy, the orbital angle can wind and its derivatives are singular. Integration by parts across that point must retain the puncture boundary. A naive calculation that drops it need not agree with the pointwise kernel. We therefore use the fully separated three-band models for the main existence claim. As an additional smooth continuation, a two-level regulator $m\sigma_y$ gives equal and opposite Berry-moment and dipole-curl contributions. This regulator breaks time reversal and is a mathematical limit, not part of the reciprocal proposal. No statement about every possible ordering of nodal, damping, and static limits is inferred from it.

For a real state the selection rule has a useful interpretation in terms of the quantum metric \cite{Provost1980}. Let $\bm X=D_x\bm u_n$, $\bm Y=D_y\bm u_n$ and write $\ell=\ii A$ with $A$ real antisymmetric. The component of $\bm Y$ parallel to $\bm X$ gives zero in $\bm X^TA\bm Y$. Cauchy's inequality for the remaining component therefore gives
\begin{equation}
 |\mathcal F_n|\le\|\ell\|
 \sqrt{\|\bm X\|^2\|\bm Y\|^2-(\bm X^T\bm Y)^2}
 =\|\ell\|\sqrt{\det\mathsf g_n}.
\end{equation}
The metric supplies an amplitude bound and a rank condition, not the sign of the response. A normalized real two-component state has a one-dimensional tangent space and $\det\mathsf g_n=0$. A third real component permits rank two, but does not guarantee either a nonzero kernel or a nonzero Brillouin-zone integral.

For $\ell=0\oplus\sigma_y$ and a normalized real three-component state,
\begin{equation}
 \mathcal F_n=\bm e_s\cdot(\partial_x\bm u_n\times\partial_y\bm u_n)
 =u_{s,n}\,\bm u_n\cdot(\partial_x\bm u_n\times\partial_y\bm u_n).
 \label{supp:eq:orientedarea}
\end{equation}
Both expressions are unchanged under $\bm u_n\mapsto-\bm u_n$. This oriented area in real mode-composition space can be nonzero while the ordinary complex-state Berry curvature vanishes. It explains the two-dimensional orbital texture needed for multimode activation.

\subsection{Relation to Euler geometry}
\label{supp:sec:euler}
The three-component area has a direct relation to real-band Euler geometry \cite{Ahn2019,Bouhon2020}. Choose an orthonormal frame $(\bm e_1,\bm e_2)$ of the plane perpendicular to a smooth real eigenvector $\bm u$, with $\bm e_1\times\bm e_2=\bm u$. The local Euler connection and curvature are
\begin{equation}
 a_i=\bm e_1\cdot\partial_i\bm e_2,\qquad
 \mathcal E=\partial_xa_y-\partial_ya_x
 =\bm u\cdot(\partial_x\bm u\times\partial_y\bm u).
\end{equation}
To see the last equality, resolve $\partial_i\bm e_{1,2}$ into this orthonormal triad. The in-plane connection cancels in $\partial_xa_y-\partial_ya_x$, leaving the determinant of the two normal components, which is the displayed scalar triple product. For a globally orientable complementary plane its Euler number is $\xi=(2\pi)^{-1}\int_{\BZ}\mathcal E\,d^2q$. The angular kernel instead obeys
\begin{equation}
 \mathcal F=u_s\mathcal E
 =\frac12(\partial_xd^\ell_y-\partial_yd^\ell_x),\qquad
 d^\ell_i=\bm e_s\cdot(\bm u\times\partial_i\bm u).
 \label{supp:eq:weightedEuler}
\end{equation}
Reversing $\bm u$ reverses the chosen plane orientation and hence $\mathcal E$, while leaving $u_s\mathcal E$ and $d^\ell_i$ unchanged. The angular observable therefore measures an orientation-independent weighted curvature. For a smooth periodic band $d^\ell_i$ is a globally defined one-form, so $\int_{\BZ}\mathcal F\,d^2q=0$. Its entropy-weighted integral can nevertheless be finite. This explains why the transport coefficient probes real-band geometry without measuring a quantized Euler class.

The minimal model illustrates the distinction. Here $h_p-\epsilon_s I$ is positive throughout the Brillouin zone, with a uniform lower bound $1.279$ meV. The lowest eigenvector therefore has a nonzero $s$ component and can be oriented globally by $u_s>0$. On the $160^2$ grid $\min u_s=0.94377$. Its image lies in a contractible hemisphere: the complementary plane is Euler trivial, although its orbital response is finite. Conversely, set
\begin{align}
 \bm v&=(1+\cos q_x+\cos q_y,\ \sin q_x,\ \sin q_y)^T,\qquad
 \bm u=\bm v/|\bm v|,\\
 P&=\bm u\bm u^T,\qquad h=14P+30(I-P)\ {\rm meV}.
\end{align}
The complementary plane has $\xi=-2$ in this orientation. The isolated $14$ meV band is flat, however, and its angular response is exactly zero because its entropy is constant. The complementary pair is degenerate; no separate isolated-band response is assigned to its members. This example rules out an Euler-class-only nonzero lower bound on the isolated-band contribution. Direct quadrature gives $\xi=-2$ and zero contribution within numerical precision, while Eq.~(\ref{supp:eq:weightedEuler}) agrees with the original interband kernel to $5.6\times10^{-17}$ in the minimal model. The selection principle uses the local curvature and its dispersion weighting, rather than requiring a topological phase.

\section{Local-mode response, activation, and dispersion controls}
\label{supp:sec:model}
In the minimal model the $p$ block is $h_p=\epsilon_p I+d_x\sigma_x+(d_z+\delta)\sigma_z$. The onsite $p_x,p_y$ energies are $E_p+\delta,E_p-\delta$, so their onsite splitting is $2\delta$. For weak $g$, the lowest eigenvector has
\begin{equation}
 \begin{pmatrix}\beta_x\\\beta_y\end{pmatrix}
 =(\epsilon_s I-h_p)^{-1}g\begin{pmatrix}\sin q_x\\\sin q_y\end{pmatrix}+O(g^3).
\end{equation}
The leading $s$-band kernel is the Jacobian of the admixture amplitudes $(\beta_x,\beta_y)$ displayed in the main text. The complete calculation retains all three bands, including the changes of the two $p$-derived eigenstates. The $g\mapsto-g$ change is a constant basis transformation $\operatorname{diag}(-1,1,1)$, so the response is even in $g$. No sign reversal under this transformation is predicted.

We sample $q_i=-\pi+(j+1/2)2\pi/N$ and evaluate analytic Hamiltonian derivatives. For separated bands all inverse gaps are used directly. A $10^{-9}$ meV threshold excludes exactly degenerate denominators; the gapped results have gaps about nine orders of magnitude larger. No phenomenological broadening enters the reported clean coefficients. Entropies are computed with \texttt{expm1} to preserve numerical accuracy.

\begin{table}[ht]
\centering
\caption{Core gapped results at $T=300$ K. Both models use $\delta=2.5$ meV and the constant local generator. Response coefficients have six significant figures; gaps have five decimal places.}
\begin{tabular}{lrrr}\toprule
Model & $N$ & $\BA$ & minimum gap (meV)\\\midrule
Minimal, $g=1$ meV & 80 & $-3.51239\times10^{-5}$ & 1.06331\\
 &160&$-3.51239\times10^{-5}$&1.06107\\
 &320&$-3.51239\times10^{-5}$&1.06051\\
Maxwell fit &80&$-8.23353\times10^{-6}$&0.97990\\
 &160&$-8.23353\times10^{-6}$&0.97723\\
 &320&$-8.23353\times10^{-6}$&0.97656\\\bottomrule
\end{tabular}
\end{table}

The weak-hybridization results are summarized by
\begin{equation}
\begin{array}{c|ccc}
 g\ (\mathrm{meV})&0.05&0.1&0.2\\\hline
 \BA&-1.17697\times10^{-7}&-4.69620\times10^{-7}&-1.86000\times10^{-6}.
\end{array}
\end{equation}
The antisymmetric intrinsic coefficient of the decoupled minimal model is below $6\times10^{-25}$. The temperature dependence is shown in main-text Fig.~1(b). The response at $g=1$ meV is approximately $-1.269\times10^{-5}$ at $50$ K, $-2.748\times10^{-5}$ at $100$ K, and $-3.608\times10^{-5}$ at $800$ K. The sign of the extremely small activated low-temperature response is not constrained to equal its room-temperature sign.

The splitting breaks fourfold symmetry. Table~\ref{supp:tab:tensors} reports both components and their antisymmetric and symmetric combinations, including the decoupled circular-cavity calibration. Their half differences reproduce the direct kernel. The minimal-model constants are $E_s=14$, $b_s=0.3$, $E_p=19.9362132696$, $b_p=0.71898$, $A=0.98492$, and $C=0.18712$ meV. These are specified model inputs; the main text rounds $E_p$ for readability. The selection rule cancels the antisymmetric intrinsic tensor; symmetric off-diagonal components can remain finite.

\begin{table}[ht]
\centering
\caption{Complete off-diagonal tensors at $T=300$ K, $\delta=2.5$ meV, and $N=160$. All four numeric columns are in units of $10^{-5}$. In the calibrated onsite model, $\lambda_{sp}$ multiplies its fitted $s$--$p$ couplings; it is distinct from the minimal model's $g$ in meV. The displayed zeros are the exact antisymmetric selection rule.}
\label{supp:tab:tensors}
\begin{tabular}{lrrrr}\toprule
Model & $\alpha^L_{yx}/\kb$ & $\alpha^L_{xy}/\kb$ & $\BA$ & $B_{\rm S}$\\\midrule
Minimal, $g=0$ & $-2.56604$ & $-2.56604$ & $0$ & $-2.56604$\\
Minimal, $g=1$ meV & $-4.63988$ & $2.38489$ & $-3.51239$ & $-1.12750$\\
Calibrated onsite, $\lambda_{sp}=0$ & $0.405025$ & $0.405025$ & $0$ & $0.405025$\\
Calibrated onsite, $\lambda_{sp}=1$ & $-3.26603$ & $-1.61933$ & $-0.823353$ & $-2.44268$\\\bottomrule
\end{tabular}
\end{table}

For a smooth periodic gapped dipole, $\int_{\BZ}\mathcal F_n\,d^2q=0$. At high temperature,
\begin{equation}
 c_1(n_n)=1+\ln(\kb T/E_n)+O[(E_n/\kb T)^2].
\end{equation}
The constant and common logarithmic pieces consequently integrate to zero. The remaining energy-dependent logarithm permits a finite high-temperature limit. This is an integrated statement about the dipole curl, not a pointwise assertion that the kernels sum to zero.

The role of dispersion follows without an additional response approximation. Since
\begin{equation}
 \frac{dc_1(n_B(E))}{dE}=-\frac{E}{(\kb T)^2}n_B(E)[1+n_B(E)]\equiv-w(E,T),
\end{equation}
integration by parts gives
\begin{equation}
 \BA=\frac12\av{\sum_n w_n
 [(\partial_xE_n)d^\ell_{n,y}-(\partial_yE_n)d^\ell_{n,x}]}.
 \label{supp:eq:weighteddipole}
\end{equation}
No occupation imbalance multiplying a diagonal band angular moment is needed. The response instead depends on the transverse correlation between group velocity and the internal angular dipole. A zero mean angular moment does not imply that this dipole vanishes: opposite local angular weights can cancel their monopole while retaining a first spatial moment.

Define a controlled deformation that keeps every eigenvector fixed,
\begin{equation}
 h_\eta(\bm q)=\sum_n[\bar E_n+\eta(E_n(\bm q)-\bar E_n)]P_n(\bm q),
 \qquad 0\le\eta\le1,
\end{equation}
where $\bar E_n$ is the mean band energy. The projectors, dipoles, and kernels are unchanged. At $\eta=0$, each entropy is constant in momentum, so $\BA=0$ exactly for the gapped model. This is a theoretical deformation; it is not claimed to preserve a short hopping range or a specific Maxwell geometry. The $160^2$ results at $300$ K are
\begin{center}
\begin{tabular}{crr}\toprule
$\eta$ & Minimal model $\BA$ & Gapped Maxwell fit $\BA$\\\midrule
0 & $-1.39\times10^{-19}$ & $-5.69\times10^{-18}$\\
0.1 & $-3.72831\times10^{-6}$ & $-2.95790\times10^{-6}$\\
0.5 & $-1.81522\times10^{-5}$ & $-1.00124\times10^{-5}$\\
1 & $-3.51239\times10^{-5}$ & $-8.23353\times10^{-6}$\\\bottomrule
\end{tabular}
\end{center}
The local kernel remains finite, with maximum magnitude about $0.0654$ in the minimal model. This distinguishes the existence of geometric texture from a nonzero integrated thermal response. Equation~(\ref{supp:eq:weighteddipole}) agrees with the direct kernel integral to $7.5\times10^{-20}$ for the minimal model and $5.6\times10^{-18}$ for the Maxwell fit.

Within the three-mode model, the formal high-temperature limit is
\begin{equation}
 \BA(\infty)=-\av{\sum_n\ln(E_n/E_*)\mathcal F_n},
\end{equation}
independent of the reference energy $E_*$. It equals $-3.62410\times10^{-5}$ for the minimal gapped model and $-9.18608\times10^{-6}$ for the gapped Maxwell fit. A continuum high-temperature prediction would additionally require the thermally occupied higher Maxwell bands, so these limits characterize the specified three-mode models.

\section{Symmetry, transverse response, and downfolding}
\label{supp:sec:symmetry}
Inversion gives $E_n(-\bm q)=E_n(\bm q)$ and $\Omega_{xy,n}(-\bm q)=\Omega_{xy,n}(\bm q)$, whereas spinless time reversal gives $\Omega_{xy,n}(-\bm q)=-\Omega_{xy,n}(\bm q)$. Their combination enforces pointwise zero ordinary Berry curvature on isolated bands. The local $L_z$ generator is inversion even and time-reversal odd, so its diagonal band expectation is zero by the same reasoning. The orbital current, being the product of two time-reversal-odd quantities, is time-reversal even. The dipole-curl kernel is inversion even and need not cancel between $\bm q$ and $-\bm q$.

The model also retains the two coordinate mirrors. Because $L_z$ is axial, its response tensor transforms under an in-plane orthogonal operation $R$ as $\alpha^L\mapsto\det(R)R\alpha^LR^T$. The mirrors set both diagonal components to zero. At $\delta=0$, fourfold rotation further gives $\alpha^L_{xy}=-\alpha^L_{yx}$. When $\delta\ne0$, write $B_{\rm S}=(\alpha^L_{yx}+\alpha^L_{xy})/(2\kb)$. For a thermal force $-\nabla T$ at angle $\phi$ to $x$, the coefficient along its counterclockwise transverse direction is
\begin{equation}
 \frac{\alpha^L_\perp(\phi)}{\kb}=\BA+B_{\rm S}\cos(2\phi).
 \label{supp:eq:angularseparation}
\end{equation}
Thus $\BA$ is the orientation-independent antisymmetric component, rather than the entire transverse response of an anisotropic sample. At $300$ K, $B_{\rm S}=-1.12750\times10^{-5}$ for the minimal gapped model and $-2.44268\times10^{-5}$ for the gapped Maxwell fit. These relations provide a theoretical way to separate the two tensor components without introducing a detector model.

Downfolding clarifies what the two-band selection rule does and does not exclude. Write $h=\left(\begin{smallmatrix}\epsilon_s&\bm t^T\\\bm t&h_p\end{smallmatrix}\right)$ and $\bm\beta=(\epsilon_sI-h_p)^{-1}\bm t=O(g)$. A frame for the $p$-like invariant subspace has the expansion
\begin{equation}
 W_p=\begin{pmatrix}-\bm\beta^T+O(g^3)\\I-\bm\beta\bm\beta^T/2+O(g^4)\end{pmatrix}.
\end{equation}
Projecting the Hamiltonian, angular generator, and position gives, to second order,
\begin{align}
 h_{p,\rm eff}&=h_p-\tfrac12(\bm t\bm\beta^T+\bm\beta\bm t^T)+O(g^4),\label{supp:eq:downfoldh}\\
 \ell_{p,\rm eff}&=\left(1-\tfrac12\bm\beta^T\bm\beta\right)\sigma_y+O(g^4),\\
 \mathcal A_{p,i}&=\frac{\ii}{2}
 (\bm\beta\partial_i\bm\beta^T-\partial_i\bm\beta\bm\beta^T)+O(g^4).
 \label{supp:eq:downfold}
\end{align}
The last two terms are of the same order as the leading orbital Nernst response. A two-band calculation that keeps only $h_{p,\rm eff}$ and resets the position and angular generator to their original constants therefore discards relevant information. The fixed-operator selection rule cannot be applied to this downfolded operator set. At three generic momenta, halving $g$ from $0.2$ to $0.1$ to $0.05$ meV reduces the residuals of Eqs.~(\ref{supp:eq:downfoldh})--(\ref{supp:eq:downfold}) by approximately a factor of 16, consistent with the fourth-order remainders. For example, the angular-matrix residual is $1.73\times10^{-5}$, $1.09\times10^{-6}$, and $6.81\times10^{-8}$.

This expansion is a statement about the projected $p$ sector. In our numerical parameters the $s$ band is thermally occupied and contributes to the total coefficient; dropping its thermal contribution would not be justified. More generally, a weakly populated additional band can still enter the interband denominators of another band. A three-band pathway does not represent sequential population of three modes.

\section{Canonical scalar field and cavity-resolved angular current}
\label{supp:sec:canonical}
The field realization defines its angular density before a frequency or mode projection. Use rescaled real canonical coordinates $\Phi_{R,a},\Pi_{R,a}$ with $[\Phi_{R,a},\Pi_{R',b}]=\ii\delta_{RR'}\delta_{ab}$, where $R$ labels cells and $a$ labels spatial mesh points. The scalar TM Drude Hamiltonian is
\begin{equation}
 \hat H=\frac{E_0}{2}(\Pi^T\Pi+\Phi^TK\Phi),\qquad
 K=-\nabla_{\rho}^2+\kappa^2\chi_{\rm host}.
\end{equation}
The physical angular momentum equals $\hbar$ times the dimensionless bilinear below. Set $D_\varphi=\rho_x\nabla_{\rho,y}-\rho_y\nabla_{\rho,x}$ and let $C_{\rm cav}$ multiply by the vacuum-cavity indicator. We define
\begin{equation}
 \hat\ell_R^{\rm cav}=\frac12(\Pi_R^TG\Phi_R+\Phi_R^TG^T\Pi_R),
 \qquad G=-C_{\rm cav}D_\varphi.
 \label{supp:eq:fielddensity}
\end{equation}
This is the Hermitian canonical angular density integrated over the vacuum cavity. In the scalar TM representation it describes the orbital rotation of the scalar vector-potential coordinate about the cavity center. It is a declared canonical observable; no equality to a total field-plus-material mechanical torque or to an angular flux through a detector is assumed.

The cavity does not touch the cell boundary. Its finite-difference derivative and window therefore define a real cell-local matrix $G$. The displacement used for transport is $\sum_R R_i\hbar\ell_R^{\rm cav}$; the angular matrix commutes with this cell position. This choice fixes both the density and the position convention. It is different from first projecting the unwindowed positive-frequency angular matrix of Sec.~\ref{supp:sec:angular} and multiplying it by a continuum position operator.

For any positive reference frequency $\omega_r$, introduce $a=(\sqrt{\omega_r}\Phi+\ii\Pi/\sqrt{\omega_r})/\sqrt2$ and $\Psi_{\bm q}=(a_{\bm q},a_{-\bm q}^\dagger)^T$. Then $\hat H=(E_0/2)\sum_{\bm q}\Psi_{\bm q}^\dagger h_B\Psi_{\bm q}$, with
\begin{equation}
 h_B=\frac12\begin{pmatrix}
 K/\omega_r+\omega_r I&K/\omega_r-\omega_r I\\
 K/\omega_r-\omega_r I&K/\omega_r+\omega_r I
 \end{pmatrix}.
\end{equation}
Writing $\Lambda=\ii(G-G^T)/2$ and $S=(G+G^T)/2$ gives the local angular vertex
\begin{equation}
 \hat\ell_R^{\rm cav}=\frac12\Psi_R^\dagger\mathcal O\Psi_R,\qquad
 \mathcal O=\begin{pmatrix}\Lambda&\ii S\\-\ii S&-\Lambda\end{pmatrix}.
 \label{supp:eq:nambuangular}
\end{equation}
For a circular window the continuum $S$ vanishes because $D_\varphi$ commutes with the window. For an ellipse it generally does not. The off-diagonal canonical blocks must then be retained even though the underlying Maxwell medium is passive.

The finite-wave-vector density is $\rho_{\mathcal O}(\bm Q)=\frac12\sum_{\bm q}\Psi_-^\dagger\mathcal O\Psi_+$, and its exact time derivative contains
\begin{equation}
 \dot\rho_{\mathcal O}(\bm Q)=\frac{\ii E_0}{2\hbar}
 \sum_{\bm q}\Psi_-^\dagger
 [h_-\Sigma_3\mathcal O-\mathcal O\Sigma_3h_+]\Psi_+,
 \quad\Sigma_3=\operatorname{diag}(I,-I).
\end{equation}
\subsection{Continuity vertices and thermal response}
Write $h_\pm=h_B(\bm q\pm\bm Q/2)$ and $u_i^B=\partial_i h_B$. In the convention in which an observable density is $\Psi^\dagger(\cdots)\Psi/2$, the dimensionless current and torque vertices are
\begin{align}
 \mathsf j_i&=\tfrac12(u_i^B\Sigma_3\mathcal O+\mathcal O\Sigma_3u_i^B),\\
 \mathsf t&=\ii(h_B\Sigma_3\mathcal O-\mathcal O\Sigma_3h_B).
 \label{supp:eq:bdgvertices}
\end{align}
Indeed, expansion of the finite-$\bm Q$ bracket gives $-\ii\mathsf t-Q_i\mathsf j_i+O(Q^2)$. The physical angular current has the corresponding factors of $E_0$, $\hbar$, and the lattice period. Absorbing the outer factor $1/2$ into the vertices above gives the $1/4$ current and $\ii/2$ torque prefactors of Ref.~\cite{Koyama2025}. The Luttinger perturbation is $\delta H_B=\{\psi,H_B\}/2$, with $H_B=E_0h_B$ and energy vertex $E_0(h_-+h_+)/2$. The current and torque therefore come from the same finite-wave-vector density as the thermal perturbation.

Let $h_Bt_{s,n}=s\Omega_n\Sigma_3t_{s,n}$, $t_{s,n}^\dagger\Sigma_3t_{s',m}=s\delta_{ss'}\delta_{nm}$. Completeness is $\sum_{s,n}s\,t_{s,n}t_{s,n}^\dagger=\Sigma_3$. Denote $o_n=t_{+,n}^\dagger\mathcal O t_{+,n}$ and
\begin{equation}
 \Omega^B_{ij,n}=-2\operatorname{Im}\langle\partial_i t_{+,n}|\Sigma_3|\partial_jt_{+,n}\rangle.
\end{equation}
The torque-multipole identity and equilibrium-circulation subtraction \cite{Xiao2021,Koyama2025} give the local-observable band kernel
\begin{equation}
 \mathscr K_{ij,n}=\Omega^B_{ji,n}o_n+\partial_jd^{\rm can}_{n,i},\qquad
 \frac{\alpha^L_{ij}}{\kb}=\av{\sum_{n>0}c_1(n_n)\mathscr K_{ij,n}}.
 \label{supp:eq:bdgresponse}
\end{equation}
Here $n>0$ denotes physical positive-frequency bands, not the signs of virtual intermediate states. In the notation of Ref.~\cite{Koyama2025}, $\mathscr K_{ij,n}$ is the geometric combination $-\Omega^{ij}_no_n+\partial_jd_{n,i}$ in its Eq.~(78), with $k_i$ converted to $q_i$. We use this geometric identity; the overall thermal sign is fixed below.

For a positive energy $E_n=E_0\Omega_n$, set $g_n=\kb T\ln(1-e^{-E_n/(\kb T)})$. The thermal-force weight following from the local grand potential is
\begin{equation}
 \mathscr W_{n,j}=-\frac{\nabla_jT}{T}(E_nn_n-g_n)
 =-\kb c_1(n_n)\nabla_jT.
 \label{supp:eq:thermalforce}
\end{equation}
Consequently $J_i^L=\av{\sum_{n>0}\mathscr K_{ij,n}\mathscr W_{n,j}}=-\alpha^L_{ij}\nabla_jT$. This fixes the sign by $-\partial_Tg_n=\kb c_1(n_n)$ and agrees with the thermoelectric convention in Eq.~(11) of Ref.~\cite{Xiao2021}. In particular, the negative sign in Eq.~(\ref{supp:eq:thermalforce}) must be retained when translating a formula written using a thermal-force symbol.

The Nambu $1/2$ removes double counting: a paired physical mode contributes $(1/2)[n_B(E)+1+n_B(E)]E=E[n_B(E)+1/2]$ to the equilibrium energy. Its temperature derivative counts one oscillator, and the temperature-independent vacuum term supplies no entropy. Thus there is no residual factor of two or one half in Eq.~(\ref{supp:eq:bdgresponse}). In the number-conserving limit, $H_B=\operatorname{diag}(h(\bm q),h^T(-\bm q))$ and $\mathcal O=\operatorname{diag}(\ell,\ell^T)$; particle--hole matrix elements between the two blocks vanish. The positive-sector dipole and Eq.~(\ref{supp:eq:bdgresponse}) then reduce to Eq.~(\ref{supp:eq:general}).

The matrix algebra uses Hermiticity, cell locality, and the bosonic particle--hole constraint $\Sigma_1\mathcal O^*\Sigma_1=\mathcal O$, all satisfied by Eq.~(\ref{supp:eq:nambuangular}); it does not require $\mathcal O$ to be real or block diagonal. There is also no residual bulk torque-monopole source. Since $(\mathsf t)_{s m,+n}=\ii(s\Omega_m-\Omega_n)\mathcal O_{s m,+n}$, completeness gives
\begin{equation}
 -2\operatorname{Im}\sum_{(s,m)\ne(+,n)}s\,
 \frac{(u_j^B)_{+n,s m}(\mathsf t)_{s m,+n}}
 {(\Omega_n-s\Omega_m)^2}=\partial_jo_n=0.
 \label{supp:eq:bdgtorquesource}
\end{equation}
Combined $PT$ makes $o_n$ and $\Omega^B_{ij,n}$ vanish for each isolated band. Equation~(\ref{supp:eq:bdgresponse}) therefore contains only the angular dipole derivative in this system.

\subsection{Constructive extension to a general local Nambu vertex}
\label{supp:sec:nambuproof}
The diagonal-observable restriction in Eqs.~(8)--(9) of Ref.~\cite{Koyama2025} cannot simply be dropped. Here we reduce our non-diagonal observable to that proven case, including the torque multipoles and the subtraction of circulating current. The argument applies on each finite spatial mesh, before taking a continuum limit.

\emph{Step 1: a local canonical reduction.} Let $\mathcal O=\mathcal O^\dagger$ obey $\Sigma_1\mathcal O^*\Sigma_1=\mathcal O$, and choose a constant $c>\|\mathcal O\|$. Write
\begin{equation}
 \mathcal O=A-cI,\qquad A=\mathcal O+cI>0.
 \label{supp:eq:positiveextension}
\end{equation}
A positive bosonic quadratic form has a canonical diagonalization \cite{Colpa1978}. In detail, diagonalize the Hermitian matrix $A^{1/2}\Sigma_3A^{1/2}$ as $U\operatorname{diag}(\lambda,-\lambda)U^\dagger$, with $\lambda_a>0$. The matrix
\begin{equation}
 B=A^{-1/2}U\operatorname{diag}(\sqrt\lambda,\sqrt\lambda)
 \quad\hbox{satisfies}\quad
 B^\dagger\Sigma_3B=\Sigma_3,\qquad
 B^\dagger AB=\operatorname{diag}(\lambda,\lambda).
 \label{supp:eq:localcanonical}
\end{equation}
Particle--hole paired columns can be chosen so that $\Sigma_1 B^*\Sigma_1=B$. The transformation $\Psi_R=B\widetilde\Psi_R$ acts only within each cell and is independent of $\bm q$, $\bm R$, temperature, and the thermal source. It preserves the cell position, introduces no position connection, and maps $A$ precisely to the real diagonal observable assumed in the reference. The second observable $cI$ already has that form. Positivity is needed only for this auxiliary construction; angular momentum itself need not be positive.

\emph{Step 2: transform the whole continuity problem.} For the $A$ problem set $h'=B^\dagger h_BB$, $A'=B^\dagger AB$, and $t'_{s,n}=B^{-1}t_{s,n}$. Since $B\Sigma_3B^\dagger=\Sigma_3$, both vertices in Eq.~(\ref{supp:eq:bdgvertices}) transform by congruence. More strongly, at finite $\bm Q$,
\begin{equation}
 h'_-\Sigma_3A'-A'\Sigma_3h'_+
 =B^\dagger(h_-\Sigma_3A-A\Sigma_3h_+)B.
 \label{supp:eq:finiteqcovariance}
\end{equation}
This retains the second-order term
$Q_iQ_j[(\partial_i\partial_jh_B)\Sigma_3A-A\Sigma_3(\partial_i\partial_jh_B)]/8$,
which generates the torque quadrupole. The energy source transforms in the same way. Thus the reduction applies to the density, displacement, torque, current, and their source derivatives together, rather than only to a dipole matrix element.

\emph{Step 3: retain the torque multipoles until circulation is removed.} To display their cancellation, use the half-vertices $\jmath_i=\mathsf j_i/2$ and $\vartheta=\mathsf t/2$, as in Ref.~\cite{Koyama2025}, and work in dimensionless frequency units. For a positive band define
\begin{align}
 C_{ij,n}&=-2\operatorname{Im}\sum_{b\ne n}s_b
       \frac{(u_j^B)_{nb}(\jmath_i)_{bn}}{(\Omega_n-\varepsilon_b)^2},\\
 d^{\jmath_i}_{n,j}&=\operatorname{Im}\sum_{b\ne n}s_b
       \frac{(u_j^B)_{nb}(\jmath_i)_{bn}}{\Omega_n-\varepsilon_b},
 \qquad \varepsilon_b=s_b\Omega_b.
\end{align}
The diagonal-observable torque identities, Eqs.~(51)--(52) of that reference, read in these conventions
\begin{align}
 q^\vartheta_{ij,n}
 &=-d^{\jmath_i}_{n,j}+\tfrac12d_{n,i}\partial_j\Omega_n+m_{ji,n},\\
 \chi^\vartheta_{ij,n}
 &=-C_{ij,n}+\tfrac12\Omega^B_{ji,n}o_n+\tfrac12\partial_jd_{n,i}.
 \label{supp:eq:torquemultipoles}
\end{align}
Here $q^\vartheta$ is the torque quadrupole, $\chi^\vartheta$ its dipole response to the drive, and $m_{ij}=-m_{ji}$ is the local circulation moment. The first identity combines the conventional-current dipole with the torque quadrupole. Together with the Berry grand-potential term it leaves an equilibrium current of the form $J^{\rm eq}_i=\partial_j M_{ij}$, with $M_{ij}=-M_{ji}$. This term is subtracted, including its temperature dependence. The remaining driven coefficient contains
\begin{equation}
 2(C_{ij,n}+\chi^\vartheta_{ij,n})
 =\Omega^B_{ji,n}o_n+\partial_jd_{n,i}.
 \label{supp:eq:multipoleclosure}
\end{equation}
The factor two here converts the half-vertex convention; it is not an additional count of thermal oscillators. Multiplying by Eq.~(\ref{supp:eq:thermalforce}) gives Eq.~(\ref{supp:eq:bdgresponse}). Keeping only a conventional-current correlator, or omitting the equilibrium subtraction, would not give this expression.

\emph{Step 4: pull back and subtract.} All quantities in the preceding step are linear in the observable vertex. The multipoles are coefficients of the same finite-$\bm Q$ density and source expansion in Eq.~(\ref{supp:eq:finiteqcovariance}); their canonical covariance therefore follows before that expansion is truncated. Explicitly,
\begin{equation}
 (t'_a)^\dagger A't'_b=t_a^\dagger At_b,\qquad
 (t'_a)^\dagger(\partial_i h')t'_b=t_a^\dagger(\partial_i h_B)t_b,
 \qquad \Omega^{B\prime}_{ij,n}=\Omega^B_{ij,n}.
\end{equation}
Consequently $o_n$, $d_{n,i}$, both torque identities, and the circulating term are unchanged by this local representation change. Apply the diagonal result to $A'$ in the transformed Hamiltonian, pull it back, and subtract the result for $cI$ in the original Hamiltonian. Linearity yields Eq.~(\ref{supp:eq:bdgresponse}) for $\mathcal O=A-cI$. Any torque-monopole source must be retained for the two auxiliary observables until they are subtracted; only the total angular observable has the pointwise zero source in Eq.~(\ref{supp:eq:bdgtorquesource}). We do not assume separate conservation of the auxiliary observables.

This proves the required extension without assuming $[\mathcal O,\Sigma_3]=0$ or discarding the $\ii S$ blocks. The proof relies on a momentum-independent local canonical transformation; it does not extend the formula to an arbitrary momentum-dependent projected observable. Independent three- and five-mode examples with nonzero anomalous vertices verify the canonical reduction, finite-$\bm Q$ continuity covariance, and dipole decomposition with maximum residual below $1.5\times10^{-13}$. These checks support the algebra above; they are distinct from the scalar-dipole reduction below.

\subsection{Exact reduction of the canonical dipole}
For $K\psi_n=\Omega_n^2\psi_n$, the signed-frequency eigenvectors of $\Sigma_3h_B$ are
\begin{equation}
 t_{s,n}=\frac12\begin{pmatrix}a_n+s b_n\\a_n-s b_n\end{pmatrix}\otimes\psi_n,
 \quad a_n=\sqrt{\omega_r/\Omega_n},\quad b_n=a_n^{-1},\quad s=\pm1.
\end{equation}
Their metric norm is $s$. The angular dipole for a positive-frequency band contains both signs as intermediate states,
\begin{equation}
 d^{\rm can}_{n,i}=\operatorname{Im}
 \sum_{(s,m)\ne(+,n)}s\,
 \frac{(t_{+,n}^\dagger\partial_i h_Bt_{s,m})
       (t_{s,m}^\dagger\mathcal O t_{+,n})}{\Omega_n-s\Omega_m}.
 \label{supp:eq:bdgdipole}
\end{equation}
The reference frequency cancels. For $m\ne n$, the two $\Lambda$ coefficients combine as
\begin{equation}
 \frac{\Omega_n+\Omega_m}{4\Omega_n\Omega_m(\Omega_n-\Omega_m)}
 -\frac{\Omega_n-\Omega_m}{4\Omega_n\Omega_m(\Omega_n+\Omega_m)}
 =\frac1{\Omega_n^2-\Omega_m^2}.
\end{equation}
The $S$ contributions cancel between signs for $m\ne n$. The remaining negative-frequency state with $m=n$ contributes $(\partial_iK)_{nn}S_{nn}/(4\Omega_n^2)$. Therefore
\begin{align}
 d^{\rm can}_{n,i}&=d^\Lambda_{n,i}+\frac{S_{nn}}{2\Omega_n}\partial_i\Omega_n,\\
 d^\Lambda_{n,i}&=\operatorname{Im}\sum_{m\ne n}
 \frac{(\partial_iK)_{nm}\Lambda_{mn}}{\Omega_n^2-\Omega_m^2}
 =\operatorname{Im}\langle D_i\psi_n|\Lambda|\psi_n\rangle.
 \label{supp:eq:canonicalreduction}
\end{align}
The extra dipole is $S_{nn}\partial_iE_n/(2E_n)$, parallel to the band velocity. Its contribution to the integrated antisymmetric response vanishes in Eq.~(\ref{supp:eq:weighteddipole}) because the cross product of the two energy derivatives is zero. For a smooth isolated band this cancellation follows pointwise in the velocity--dipole form; no small-frequency or weak-window approximation is made. Thus the field antisymmetric coefficient uses the $\Lambda$ dipole curl and, in the real $PT$ frame,
\begin{equation}
 \mathcal F_n^{\rm field}=-\operatorname{Im}
 \langle D_x\psi_n|\Lambda|D_y\psi_n\rangle.
\end{equation}
This includes the virtual canonical frequency sectors; it is not obtained by inserting a number-preserving frequency-weighted angular matrix into the original onsite formula. Independent finite-dimensional calculations of Eq.~(\ref{supp:eq:bdgdipole}) agree with Eq.~(\ref{supp:eq:canonicalreduction}) to $3.4\times10^{-16}$ over three reference frequencies. Including a nonzero symmetric $S$ verifies the extended identity to $4.9\times10^{-17}$ and its vanishing transverse velocity--dipole term to $5.7\times10^{-18}$.

\section{Low-temperature field response}
\label{supp:sec:activation}
The field calculation uses elliptical vacuum cavities with semiaxes $(R_x,R_y)=(1,0.65)$, square period $2.2$, and $\kappa=4$. The energy unit is $E_0\simeq6.682$ meV, fixed by the isolated circular-cavity reference in Sec.~\ref{supp:sec:calibration}. This is a new directly solved geometry; no mass term is added to a fitted Hamiltonian. The angular window is the same vacuum region, integrated with $6\times6$ subpixels per mesh pixel.

The cell Bloch gauge puts phases on links crossing the cell boundary. Thus $\Lambda$ and $S$ are constant intracell matrices, and $K(\bm q)$ includes the Bloch phase. We compute scalar eigenvectors on the complete $40^2$ or $48^2$ spatial mesh. Derivatives use central differences with parallel phase alignment and removal of the diagonal band connection. They are full spatial vectors, so their virtual intermediate content is not limited to the retained thermal eigenstates. Thirty-two positive-frequency bands enter the reported thermal integral. This distinguishes a thermal cutoff from a virtual-mode projection.

For a real scalar elliptic operator, the lowest band has its minimum at $\Gamma$; the discrete connected Laplacian has the corresponding positive ground-state property. Here the minimum is unique, its Hessian is positive, and the lowest band is separated from the next band. Write
\begin{equation}
 E(\bm q)=E_{\min}+\frac12\bm q^TM\bm q+O(q^4),\qquad
 \mathcal F_0(\bm q)=\mathcal F_0(\Gamma)+O(q^2).
\end{equation}
Here $M_{ij}=\partial_i\partial_jE|_{\Gamma}$ is the positive energy Hessian in dimensionless momentum coordinates, with units of meV. Since $c_1(n_B(E))=(1+E/\theta)e^{-E/\theta}+O[(1+E/\theta)e^{-2E/\theta}]$, with $\theta=\kb T$, Gaussian integration gives
\begin{equation}
 \BA(T)=\frac{E_{\min}\mathcal F_0(\Gamma)}{2\pi\sqrt{\det M}}
 e^{-E_{\min}/\theta}\,[1+O(\theta/E_{\rm var})]
 +\text{higher-band activated terms}.
 \label{supp:eq:laplace}
\end{equation}
Here $E_{\rm var}$ denotes the energy scale over which the minimum and kernel depart from their local expansions. The asymptotic formula requires temperatures small on the scale of the band dispersion, not merely smaller than $E_{\min}$. Higher thermal bands have a larger activation energy. Their virtual admixture is already present in $\mathcal F_0(\Gamma)$. A nonzero prefactor supplies an existence criterion for this field current without requiring a populated third band.

\begin{center}
\begin{tabular}{lrr}\toprule
Quantity & $40^2$ spatial mesh & $48^2$ spatial mesh\\\midrule
$E_{\min}$ (meV) & 14.5500372 & 14.5487270\\
$M_{xx}$ (meV) & 0.4109605 & 0.4112417\\
$M_{yy}$ (meV) & 0.1505572 & 0.1502118\\
$\mathcal F_0(\Gamma)$ & $1.4085971\times10^{-3}$ & $1.4067346\times10^{-3}$\\
$C_0=E_{\min}\mathcal F_0/(2\pi\sqrt{\det M})$ & 0.01311355 & 0.01310559\\\bottomrule
\end{tabular}
\end{center}
The Hessian is diagonal by the coordinate mirrors. Fourth-order energy differences at steps $0.02$ and $0.01$ agree to better than $2\times10^{-9}$ meV. The kernel changes by approximately $10^{-11}$ when its derivative step is reduced from $2\times10^{-4}$ to $10^{-4}$. Ordinary Berry curvature of the lowest state at $\Gamma$ is below $8\times10^{-14}$ in the primary calculation.

The three lowest $\Gamma$ modes have mirror parities $(+,+)$, $(-,+)$, and $(+,-)$, identifying $s,p_x,p_y$. On the $40^2$ mesh their energies are $14.5500$, $23.4991$, and $25.6611$ meV. Projecting the lowest-band derivative vectors onto the two lowest modes gives a zero kernel. Restricting them to all three gives $2.1981777\times10^{-3}$; using the complete spatial derivatives gives $1.4085971\times10^{-3}$. The higher virtual channels reduce this pointwise kernel by $35.9\%$ relative to the three-mode restriction. This comparison holds the field eigenstates and the physical angular density fixed; it is not a comparison of two independently fitted transport models.

The mirror channels give a more specific design criterion than counting modes. At $\Gamma$, $\partial_xK$ couples an $(+,+)$ ground state to $(-,+)$ states, whereas $\partial_yK$ couples it to $(+,-)$ states. The axial angular vertex links those two sectors. A third mode with $(+,+)$ parity cannot supply the missing channel. We confirm this in a second directly solved array with $\kappa=6$, $(a_x,a_y)=(2.2,1.5)$, and the same ellipse. Its first four $\Gamma$ modes have parities $(+,+),(-,+),(+,+),(+,-)$. The three-mode kernel is below $4\times10^{-29}$. Adding the fourth mode gives $3.08273\times10^{-3}$, and the complete spatial basis gives $2.76326\times10^{-3}$. Thus the first three eigenvalues are not a universal sufficient set; the required odd channel can occur higher in the spectrum. This calculation concerns the kernel at an isolated lowest-band minimum and is not used to claim a converged high-temperature coefficient for the second geometry.

A separate low-momentum integral uses 12-point Gauss--Hermite quadrature in each direction with the exact numerically calculated energy and kernel. Removing the activation before summing avoids numerical underflow. It gives $\BA e^{E_{\min}/(\kb T)}=0.0129823$ at $0.05$ K, $0.0130875$ at $0.01$ K, and $0.0131110$ at $0.001$ K, approaching the predicted $C_0=0.0131135$. These scaled values are a test of the asymptotic law, not the magnitude of the raw response at those very low temperatures.

\begin{figure}[ht]
 \centering
 \includegraphics[width=0.60\textwidth]{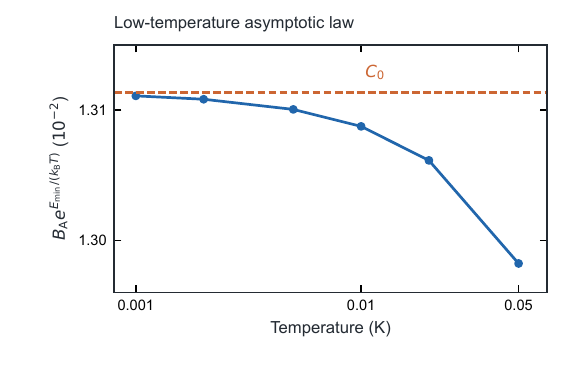}
 \caption{Independent verification of the low-temperature asymptotic law. The activation factor is removed before numerical summation, and the result approaches $C_0=0.0131135$. This ordinate is not the raw response. The unscaled field coefficient over $4$--$10$ K is shown in the main text.}
 \label{supp:fig:activationcheck}
\end{figure}

Direct Brillouin-zone integration at $10$ K gives
\begin{center}
\begin{tabular}{crr}\toprule
Spatial mesh & Momentum grid & $\BA$\\\midrule
$40^2$ & $24^2$ & $2.3181080\times10^{-11}$\\
$40^2$ & $64^2$ & $2.3180172\times10^{-11}$\\
$48^2$ & $32^2$ & $2.3151273\times10^{-11}$\\\bottomrule
\end{tabular}
\end{center}
The $40^2$--$48^2$ difference is approximately $0.13\%$. On the $40^2$, $64^2$ calculation, the lowest-band contribution is $2.310480698\times10^{-11}$, or $99.6749\%$ of the sum. The velocity--dipole expression gives $2.310480702\times10^{-11}$ for that band, a relative difference of $1.7\times10^{-9}$. For the 32-band sum the two expressions differ by $0.017\%$, below the spatial-mesh variation; the extra mismatch lies in the subleading higher-band contribution. Accordingly, the full $10$ K coefficient is quoted to three significant figures in the main text, while the isolated-band asymptotic law supplies the existence criterion. Setting every intercell link to zero removes the momentum dependence and gives zero response. The low-temperature field coefficient is therefore nonzero but small. At higher temperatures our preliminary full-band sums show substantial contributions from propagating bands with crossings, and the pointwise-kernel and integrated-dipole evaluations no longer agree. We do not use those unresolved high-temperature values as a converged total transport coefficient or as an enhancement claim. The isolated-lowest-band asymptotic result and the low-temperature numerical signal are the conclusions supported here.

\subsection{Cumulative virtual modes and signed mode-pair contributions}
\label{supp:sec:virtualsum}
Main-text Fig.~3(b) resolves how the angular-dipole kernel is assembled. We diagonalize the complete $40\times40$ spatial matrix at $\Gamma$, including all 1600 scalar eigenmodes, and differentiate the Bloch boundary links analytically. For the isolated ground mode,
\begin{equation}
 |D_i\psi_0\rangle=\sum_{m>0}|\psi_m\rangle c_{i,m},\qquad
 c_{i,m}=\frac{\langle\psi_m|\partial_iK|\psi_0\rangle}
 {\Omega_0^2-\Omega_m^2}.
 \label{supp:eq:derivativechannels}
\end{equation}
With $P_N$ projecting onto the $N$ lowest spatial modes, including the ground mode, define
\begin{align}
 \mathcal F_0^{(N)}&=-\operatorname{Im}
 \langle P_ND_x\psi_0|\Lambda|P_ND_y\psi_0\rangle
 =\sum_{1\le m,l<N}C_{ml},\label{supp:eq:cumulativekernel}\\
 C_{ml}&=-\operatorname{Im}[c_{x,m}^*\Lambda_{ml}c_{y,l}].
 \label{supp:eq:pairkernel}
\end{align}
Here mode indices start at zero in the equations. The figures label the lowest mode by 1. This is a nested projection of the virtual derivative vectors for one fixed field Hamiltonian and one fixed angular observable. The thermal-band count remains a separate cutoff. Intermediate degeneracies cause no pole in Eq.~(\ref{supp:eq:derivativechannels}) because the reference ground mode is isolated; the complete sum over a degenerate intermediate subspace is basis independent.

\begin{figure}[ht]
 \centering
 \includegraphics[width=0.93\textwidth]{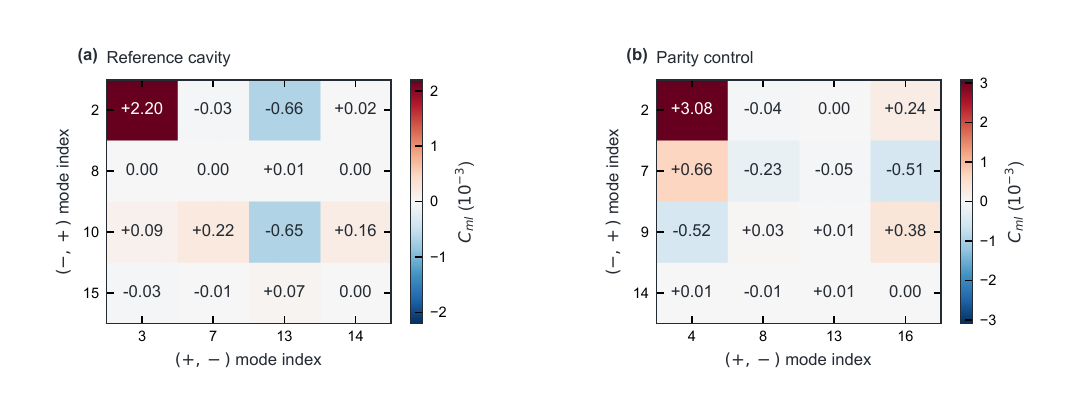}
 \caption{Signed mode-pair contributions for the reference (a) and parity-control (b) cavities of main-text Fig.~3(b). Rows and columns select the first four $(-,+)$ and $(+,-)$ modes among the lowest 20, indexed in the full energy-ordered spectrum starting at 1. Color and labels give $10^3C_{ml}$; only labels are rounded to two decimals. Higher channels can reinforce or reduce the first allowed contribution. The main text shows the complete cumulative sums.}
 \label{supp:fig:modepairs}
\end{figure}

In the reference geometry, $N=2$ gives zero, $N=3$ gives $2.19818\times10^{-3}$, and $N=1600$ gives $1.40860\times10^{-3}$. In the parity control, $N=3$ gives $1.32\times10^{-30}$, $N=4$ gives $3.08273\times10^{-3}$, and $N=1600$ gives $2.76326\times10^{-3}$. Figure~\ref{supp:fig:modepairs} resolves the first four channels in each required odd sector. Their signed matrix elements explain why adding higher modes can reduce the response after its initial activation. The cumulative curves retain all modes; the small heatmaps display selected contributions rather than an exhaustive decomposition.

The complete double sum agrees with direct contraction of the derivative vectors to an absolute error below $2.4\times10^{-18}$. Its full-basis kernels agree with the phase-aligned finite-difference derivatives used for the original field calculation to a relative difference below $2.4\times10^{-9}$. The cumulative values at $N=64$ still differ from their full-basis values by several percent, so the complete sum is used for the quoted endpoint.

\section{Band structure and scalar Maxwell calibration}
\label{supp:sec:calibration}
The ideal lossless Drude dielectric is $\epsilon(\omega,\bm r)=1-\omega_p^2\chi_{\rm host}(\bm r)/\omega^2$. For the scalar TM sector the eigenproblem can be written
\begin{equation}
 K(\bm q)\psi_n=\Omega_n^2\psi_n,\quad
 K=-\mathscr D_x^2-\mathscr D_y^2+\kappa^2\chi_{\rm host},\quad
 \mathscr D_i=\nabla_{\rho,i}+\ii q_i/a,
\end{equation}
where lengths are in units of $R$, $a=2.2$, $\kappa=\omega_pR/c=4$, and $\Omega=\omega R/c$. The positive-frequency scalar representation is $h_{\rm full}=E_0\sqrt K$, with $E_0=\hbar c/R$.

The energy unit is set by an isolated circular vacuum cavity of radius $R$ in the same infinite host. Its lowest $p$ mode has azimuthal order $|m|=1$. The radial scalar field is proportional to $J_1(\Omega r/R)$ inside and $K_1(\zeta r/R)$ outside, where $\zeta=\sqrt{\kappa^2-\Omega^2}$. Continuity of the field and its normal derivative at $r=R$ gives
\begin{equation}
 \frac{\Omega J_1'(\Omega)}{J_1(\Omega)}
 =\frac{\zeta K_1'(\zeta)}{K_1(\zeta)},\qquad
 \zeta=\sqrt{\kappa^2-\Omega^2}.
 \label{supp:eq:isolatedroot}
\end{equation}
Here $J_1$ and $K_1$ are the Bessel and modified Bessel functions; primes differentiate their arguments. At $\kappa=4$ the lowest $p$ root is $\Omega_p^{\rm circ}\simeq2.99316$. The subscript $p$ labels the spatial mode, whereas $\omega_p$ above denotes the host plasma frequency. Choosing the reference mode energy $E_p^{\rm ref}=20$ meV sets
\begin{equation}
 E_0=\frac{E_p^{\rm ref}}{\Omega_p^{\rm circ}}
 \simeq6.68191\ \mathrm{meV},\qquad E_n=E_0\Omega_n.
 \label{supp:eq:energyscale}
\end{equation}
The numerical normalization uses $\Omega_p^{\rm circ}=2.9931576706$. This is a computed boundary-matching root, not a fitted material parameter. The scale corresponds to $R\simeq29.5\,\mu$m and $\hbar\omega_p\simeq26.7$ meV in this ideal model; the isolated $s$ energy is $12.7417$ meV. All elliptical arrays retain the same $E_0$, identifying the reference $R$ with $R_x$; their eigenenergies are calculated rather than fixed to $20$ meV. The minimal model's onsite $E_p$ is a different, calibration-derived matrix element. This choice specifies a physical scale, not a particular measured Drude material or its losses.

\begin{figure}[!ht]
 \centering
 \includegraphics[width=0.97\textwidth]{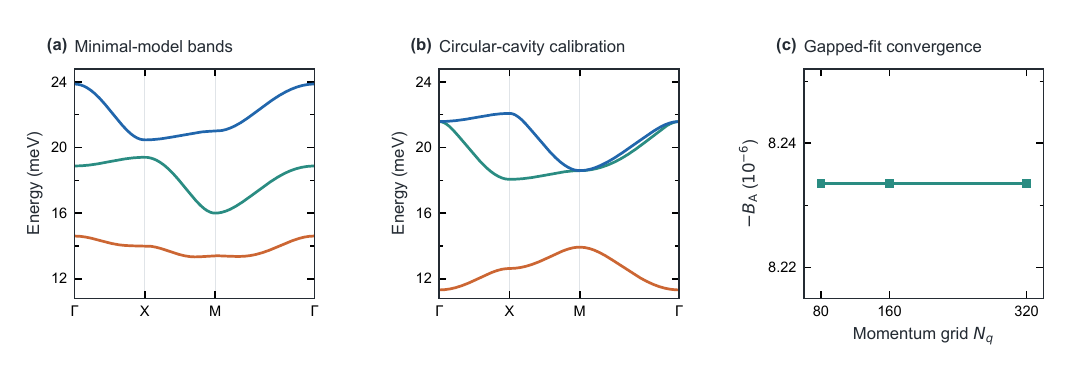}
 \caption{Band structure and calibration underlying the local-mode results. (a) Separated bands of the minimal model used in main-text Figs.~1--2, with $g=1$ meV and $\delta=2.5$ meV. (b) Bands of the circular-cavity three-mode fit before an imposed splitting. (c) Its gapped deformation with $\delta=2.5$ meV has a converged antisymmetric coefficient at $300$ K. Both models use the fixed local angular generator.}
 \label{supp:fig:calibration}
\end{figure}

A cell-periodic finite-difference grid has $40\times40$ points per cell, spacing $0.055R$, and $6\times6$ subpixels per cell pixel for the circular boundary. Uniform link phases distribute the Bloch phase across the grid. Bloch sums of isolated scalar $s,p_x,p_y$ functions include image cells from $-3$ to $3$ in each direction and are symmetrically orthonormalized to form $B$. If $V$ contains the three selected Maxwell eigenvectors, write $B^\dagger V=U\Sigma V_0^\dagger$ and define
\begin{equation}
 R_0=UV_0^\dagger,\qquad G=\operatorname{diag}(\ii,1,1),\qquad
 W=VR_0^\dagger G,\qquad h_P=W^\dagger h_{\rm full}W.
\end{equation}
This anchors the orbital frame rather than relying on independent eigenvector phases. The minimum overlap singular value is $0.97197$ on the $24^2$ training grid. A Fourier interpolation over $|R_x|,|R_y|\le5$ cells gives a dispersive Hamiltonian relative RMS error $9.8851\times10^{-4}$.

Using $h_P$ with the ideal local $\ell$ gives the gapped mode-model coefficient in Fig.~\ref{supp:fig:calibration}(c). At zero splitting, $\BA=-4.67807\times10^{-4}$, $-4.51956\times10^{-4}$, and $-4.47983\times10^{-4}$ on $256^2$, $512^2$, and $1024^2$ grids. These values retain their stated grid dependence. The gapped coefficient is stable when the Fourier range is changed: $-8.22508\times10^{-6}$ for range four and $-8.23306\times10^{-6}$ for range six, compared with $-8.23353\times10^{-6}$ for range five. Switching off all $s$--$p$ hoppings in the gapped fit gives a coefficient below $6\times10^{-27}$.

\section{Projected position and physical velocity}
\label{supp:sec:position}
Sections~\ref{supp:sec:position}--\ref{supp:sec:covariant} examine an unwindowed positive-frequency angular observable and a projected continuum position. This choice differs from the cavity-resolved canonical density and cell position in Sec.~\ref{supp:sec:canonical}. The identities explain why a band fit alone cannot determine a field current. Optical angular and position operators must be specified consistently \cite{Bliokh2010,Bliokh2017}.

For an isolated invariant subspace with orthonormal frame $W(\bm q)$, a continuum wave packet is represented by $Wf$. Projecting the position action gives
\begin{equation}
 W^\dagger\ii\partial_i(Wf)=\ii\partial_i f+\mathcal A_if,
 \qquad \mathcal A_i=\ii W^\dagger\partial_iW.
 \label{supp:eq:position}
\end{equation}
The frame satisfies $h_{\rm full}W=Wh_P$. Differentiating $h_P=W^\dagger h_{\rm full}W$ and using $W^\dagger\partial_iW=-\ii\mathcal A_i$ gives
\begin{equation}
 \partial_i h_P=W^\dagger(\partial_i h_{\rm full})W
                   +\ii[\mathcal A_i,h_P],\qquad
 u_i^P=\partial_i h_P-\ii[\mathcal A_i,h_P].
 \label{supp:eq:covvelocity}
\end{equation}
This is exact for the selected invariant subspace. It is the same position-connection structure that occurs in orbital-dependent Wannier interpolation \cite{Go2024}. Under $W\mapsto WU(\bm q)$,
\begin{equation}
 h_P\mapsto U^\dagger h_PU,\qquad
 \mathcal A_i\mapsto U^\dagger\mathcal A_iU+\ii U^\dagger\partial_iU,
\end{equation}
and the complete velocity transforms covariantly. The norm of the connection term alone depends on the anchored frame and is not a gauge-invariant observable.

We evaluate the left-hand side of Eq.~(\ref{supp:eq:covvelocity}) independently through the Fr\'echet derivative of the square root. In Maxwell eigenstates,
\begin{equation}
 (u_i^{\rm phys})_{nm}
 =E_0\frac{\langle\psi_n|\partial_iK|\psi_m\rangle}{\Omega_n+\Omega_m}.
 \label{supp:eq:frechet}
\end{equation}
The eigenstates are then rotated into $W$. Central differences of the anchored $W$ and $h_P$ supply $\mathcal A_i$ and $\partial_i h_P$ without a fit derivative. This compares two independent constructions of the same operator.

The calculation uses a $12^2$ shifted momentum grid, step $\Delta q=2\times10^{-4}$, and 12 retained Maxwell eigenstates. Define $\|X\|_{\rm all}^2=\sum_{\bm q,i,a,b}|X_{i,ab}(\bm q)|^2$. The two central results are
\begin{align}
 \frac{\|\partial h_P-\ii[\mathcal A,h_P]-u^P\|_{\rm all}}{\|u^P\|_{\rm all}}
 &=7.4088\times10^{-9},\\
 \frac{\|\partial h_P-u^P\|_{\rm all}}{\|u^P\|_{\rm all}}&=0.995204.
\end{align}
The elementwise RMS connection is $0.09444$ in lattice-period units and the RMS velocity correction is $0.76457$ meV in $u_i=\hbar v_i/a$ units. These are not small on the scale of the projected velocity. They are also not relative corrections to the Nernst coefficient: thermal weights, interband denominators, angular vertices, and torque terms are not contained in a velocity norm.

At the three generic points $(0.371,0.829)$, $(1.11,-0.62)$, and $(-2.24,1.83)$, reducing the finite-difference step from $10^{-3}$ to $2\times10^{-4}$ changes the relative identity residual from $2.420\times10^{-7}$ to $9.694\times10^{-9}$, as expected for a second-order difference. Increasing the retained eigenspace from 12 to 24 modes at those points gives $9.679\times10^{-9}$ and leaves the selected-subspace velocity correction unchanged. No continuum transport coefficient is extracted by replacing only the velocity vertex in the onsite formula.

\section{Angular operators and projection corrections}
\label{supp:sec:angular}
The canonical cell-centered scalar rotation is $\Lambda_z=-\ii(x\mathscr D_y-y\mathscr D_x)$ with $x,y$ measured from the cavity center and $\mathscr D_i$ the spatial Bloch derivative in Sec.~\ref{supp:sec:calibration}. The number-preserving angular matrix used in this calculation is
\begin{equation}
 (\ell^{\rm phys})_{nm}
 =\frac{\Omega_n+\Omega_m}{2\sqrt{\Omega_n\Omega_m}}
   \langle\psi_n|\Lambda_z|\psi_m\rangle.
 \label{supp:eq:angular}
\end{equation}
The symmetric frequency factor follows from the positive-frequency scalar coordinate--momentum bilinear. We use this declared canonical intracell observable; we do not identify it here with a detector's mechanical torque or with the sum of every field and material angular contribution.

Let $\ell_P=W^\dagger\ell^{\rm phys}W$. The projected position commutator is
\begin{equation}
 [r_{P,i}/a,\ell_P]=\ii\mathcal D_i\ell_P,\qquad
 \mathcal D_i\ell_P=\partial_i\ell_P-\ii[\mathcal A_i,\ell_P].
\end{equation}
On the $12^2$ grid its elementwise RMS is $0.08371$, which is nonzero. The local mode generator, by contrast, commutes with the specified cell position. This is an operator-definition difference, not only a calibration of the magnitude of angular momentum. The relative matrix difference between $\ell_P$ and $0\oplus\sigma_y$ is $0.17132$; the earlier $24^2$ grid gives $0.17132$ as well, and the mean $p$-sector angular matrix element is $0.86236$ of the ideal value.

Projecting products also produces terms outside the three-mode subspace. For $P=WW^\dagger$ and $Q=1-P$,
\begin{align}
 P\{\ell,u_i\}P-\{P\ell P,Pu_iP\}
 &=P\ell Q u_iP+Pu_iQ\ell P,\\
 [Pr_iP,P\ell P]
 &=P[r_i,\ell]P-Pr_iQ\ell P+P\ell Qr_iP.
\end{align}
Thus projecting the two factors separately need not reproduce a projected current. We compare these products with a finite higher-mode complement. With 12 retained modes, the relative difference of the conventional current products is $0.1563$ over the grid. At the three generic points it changes from $0.1409$ to $0.1434$ when 12 modes are replaced by 24. These diagnostics show a finite complement contribution; they are not claimed as a converged infinite-band current.

For completeness, the ordinary orbital-density form-factor dipole evaluated with physical matrices is
\begin{equation}
 d^{\ell,\mathrm{phys}}_{n,i}
 =\operatorname{Im}\sum_{m\ne n}
 \frac{(u_i^{\mathrm{phys}})_{nm}(\ell^{\mathrm{phys}})_{mn}}{E_n-E_m}.
\end{equation}
For the three selected bands, including the additional nine intermediate states changes its aggregate grid norm by $2.73\%$ relative to the 12-mode result. At the generic points the corresponding changes are about $22\%$ with either 12 or 24 retained modes, demonstrating the momentum dependence. The dipole alone is insufficient to infer the continuum torque-corrected transport when $[r_i,\ell]\ne0$.

The product correction has a direct interpretation for the angular displacement itself. Write $\bar r_i=r_i/a$ and define its dimensionless projection by
\begin{align}
 \mathscr D_{i,P}^{\ell}
 &=\frac12P\{\bar r_i,\ell\}P
 =\frac12\{\bar r_{i,P},\ell_P\}+C_i,\\
 C_i&=\frac12(P\bar r_iQ\ell P+P\ell Q\bar r_iP).
 \label{supp:eq:displacementprojection}
\end{align}
All terms are understood as operators acting within $P$. The physical angular displacement is $\hbar a\mathscr D_{i,P}^{\ell}$. For an invariant spectral subspace, $[P,h_{\rm full}]=0$, projecting its time derivative is equivalent to taking the commutator of this complete projected displacement with $h_P$. Thus the projected pair $(r_P,\ell_P)$ alone does not determine the projection of the full displacement current.

In the full-band energy basis, $(\bar r_i)_{nm}=-\ii(u_i^{\rm phys})_{nm}/(E_n-E_m)$ for $n\ne m$. Hence, for a selected band,
\begin{equation}
 (C_i)_{nn}=\operatorname{Im}\sum_{m\notin P}
 \frac{(u_i^{\rm phys})_{nm}\ell^{\rm phys}_{mn}}{E_n-E_m}
 =d^{\ell,\rm full}_{n,i}-d^{\ell,P}_{n,i}.
 \label{supp:eq:complementdipole}
\end{equation}
The density-dipole diagnostics above therefore measure the diagonal part of a missing angular-displacement operator. They do not, by themselves, give a correction to the dc coefficient: a complete response calculation must combine displacement, torque, thermal contact terms, and the equilibrium subtraction before deciding which terms survive.

\section{Covariant dipole curl for projected operators}
\label{supp:sec:covariant}
A kinematic identity makes the operator issue more explicit. In a momentum-dependent projected frame define
\begin{align}
 \nabla_i&=\partial_i-\ii\mathcal A_i,&
 \mathcal R_{xy}&=\partial_x\mathcal A_y-\partial_y\mathcal A_x
                  -\ii[\mathcal A_x,\mathcal A_y],\\
 |D_i c_n\rangle&=(1-|c_n\rangle\langle c_n|)\nabla_i|c_n\rangle,&
 \mathcal D_i L&=\partial_i L-\ii[\mathcal A_i,L],
\end{align}
where $c_n$ is a normalized band eigenvector and $L(\bm q)$ is a general Hermitian, dimensionless observable. Define $d_{n,i}=\operatorname{Im}\langle D_i c_n|L|c_n\rangle$, $L_{nn}=\langle c_n|L|c_n\rangle$, and the physical band connection $a_{n,i}=\ii\langle c_n|\nabla_i c_n\rangle$. Its curvature is
\begin{equation}
 \Omega_n=\partial_xa_{n,y}-\partial_ya_{n,x}
 =\langle\mathcal R_{xy}\rangle_n
       -2\operatorname{Im}\langle D_xc_n|D_yc_n\rangle.
\end{equation}
The exact dipole-curl identity is
\begin{align}
 \frac12(\partial_xd_{n,y}-\partial_yd_{n,x})
 ={}&-\operatorname{Im}\langle D_xc_n|L|D_yc_n\rangle\nonumber\\
 &+\frac12\operatorname{Im}\bigl[
 \langle D_yc_n|\mathcal D_xL|c_n\rangle
 -\langle D_xc_n|\mathcal D_yL|c_n\rangle\bigr]\nonumber\\
 &+\frac14\langle\{\mathcal R_{xy},L\}\rangle_n
 -\frac12\Omega_n L_{nn}.
 \label{supp:eq:covcurl}
\end{align}
For a short derivation, choose the band phase so that $a_{n,i}=0$ at the evaluation point. In $d_{n,i}=\operatorname{Im}\langle\nabla_i c_n|L|c_n\rangle-a_{n,i}L_{nn}$, the antisymmetric second derivative gives $[\nabla_x,\nabla_y]=-\ii\mathcal R_{xy}$; differentiating $L$ gives the second line of Eq.~(\ref{supp:eq:covcurl}); differentiating the last term gives $-\Omega_nL_{nn}/2$. All resulting terms are gauge invariant under a change of band phase and covariant under a change of projected frame, so the identity holds in any smooth gauge.

For a constant local observable with $\mathcal A_i=0$, the derivative and frame-curvature terms vanish. For the real model, $\Omega_n=L_{nn}=0$ as well, recovering the kernel of the main text. A projected angular operator generally has $\mathcal D_iL\ne0$ and a projected position can have $\mathcal R_{xy}\ne0$. Their extra terms are therefore present already at the level of dipole geometry. In particular, a momentum-dependent two-band angular operator need not obey the fixed-generator cancellation. A separate finite-difference check using a complex three-component state, a nonzero connection curvature, and a momentum-dependent Hermitian observable verifies Eq.~(\ref{supp:eq:covcurl}) to $1.7\times10^{-10}$ at step $2\times10^{-4}$. This checks the algebra, not a Maxwell transport coefficient.

Equation~(\ref{supp:eq:covcurl}) is a geometric identity for projected operators. To turn it into a transport formula for the unwindowed observable, one must also retain the complement displacement $C_i$ in Eq.~(\ref{supp:eq:displacementprojection}) and rederive the torque and equilibrium-circulation terms. We do not identify that observable with the cell-resolved canonical current established in Sec.~\ref{supp:sec:canonical}. The latter is defined before projection and uses its own continuity vertices. These projection results delimit the fixed-local-generator selection rule; they are not missing contributions to Eq.~(\ref{supp:eq:bdgresponse}).

\section{Collisionless realization and thermal scale}
\label{supp:sec:thermalscope}
\subsection{Frequency, relaxation, and length limits}
The clean intrinsic coefficient is the adiabatic interband coefficient after torque and equilibrium-circulation corrections. Its order of limits is
\begin{equation}
 \alpha^{L,\rm int}_{ij}
 =\lim_{\omega_{\rm d}\to0}\lim_{\gamma\to0^+}\lim_{Q\to0}
       \alpha^{L,\rm tr}_{ij}(Q,\omega_{\rm d};\gamma),
 \qquad
 Qv_{\max},\ \gamma/\hbar\ll|\omega_{\rm d}|\ll\Delta_{\rm dir}/\hbar.
 \label{supp:eq:cleanlimits}
\end{equation}
Here $Q$ is a physical wave number, $\omega_{\rm d}$ the thermal-drive frequency, $\gamma$ the relaxation energy, $v_{\max}$ the maximum relevant band speed, and $\Delta_{\rm dir}$ the smallest relevant direct interband gap. The system size is taken larger than the drive length before the bulk limit. Equation~(\ref{supp:eq:cleanlimits}) specifies a frequency window, not a claim that a lossless system relaxes to a stationary temperature profile. A Luttinger energy source defines the retarded coefficient; equilibrium circulation is removed before identifying its thermal-force convention. A locally equilibrated bath whose temperature oscillates need not realize this collisionless limit.

An equivalent time-domain window is $\hbar/\Delta_{\rm dir}\ll t\ll\min(\tau,(Qv_{\max})^{-1})$, where $\tau=\hbar/\gamma$. Preparation and measurement can be separated: prepare a slowly varying thermal population, decouple the preparation baths, and examine the initial collisionless evolution. The next subsection gives a minimal wave-packet implementation of this prescription and its angular-dipole readout.

In contrast, taking $\omega_{\rm d}\to0$ at fixed scattering rate asks for a stationary response and requires the corresponding collision vertex. Xiao and Niu show a cancellation of the intrinsic dipole term by a scattering contribution in their electronic weak-disorder problem \cite{Xiao2021}. This is a concrete reason not to identify the two limits. It neither proves nor excludes the same cancellation for a photonic bath. Our result is the collisionless intrinsic coefficient and its mode-selection rule; we do not postulate a universal dissipative dc coefficient.

\subsection{A prepared thermal grating and its dipole-polarization readout}
\label{supp:sec:packetprotocol}
For the real, isolated bands of the minimal model, prepare an incoherent ensemble of narrow band wave packets with local Bose populations
\begin{equation}
 f_n(\bm q,x,0)=n_n+\delta T\,\partial_T n_n\cos(Qx),
 \qquad |\delta T|\ll T_0.
\end{equation}
The packets are broad compared with a cell and narrow compared with $Q^{-1}$. After release, the leading-gradient collisionless kinetic equation has the explicit solution
\begin{equation}
 f_n(\bm q,x,t)=n_n+\delta T\,\partial_T n_n
       \cos[Q(x-v_{n,x}t)],\qquad
 v_{n,i}=\frac{a}{\hbar}\partial_i E_n.
 \label{supp:eq:ballisticgrating}
\end{equation}
The wave-packet multipole construction \cite{Culcer2004} assigns each packet zero angular monopole but the first angular moment $\hbar a\,d^\ell_{n,i}$. Define the coarse-grained angular-dipole polarization per area by
\begin{equation}
 P_i^L(\bm r,t)=\frac{\hbar}{a}\av{\sum_n d^\ell_{n,i} f_n(\bm q,\bm r,t)}.
 \label{supp:eq:packetpolarization}
\end{equation}
This is a first moment of the packet's angular-density distribution, rather than a diagonal angular population. It can be reconstructed from the angular density of resolved wave packets, or from their equivalent field correlations. For a direct example, Fourier transform a normalized Gaussian packet in band 1, centered at $\bm q=(0.371,0.829)$ with momentum standard deviation $0.12$. On $128^2$ cells its onsite angular density is $\ell_R=W_R^\dagger\ell W_R$. The total is below $5\times10^{-19}$, whereas
\begin{equation}
 \sum_R(R_x/a,R_y/a)\ell_R=(-0.01760864,\ 0.00801282).
\end{equation}
These first moments agree with the packet-weighted band dipole to $1.1\times10^{-17}$. Main-text Fig.~1(a) displays the opposite angular-density lobes of this actual lattice packet.

For the thermal ensemble, equilibrium gives $P_i^L=0$ by inversion. At the maximum initial gradient, $Qx=\pi/2$, Eqs.~(\ref{supp:eq:ballisticgrating})--(\ref{supp:eq:packetpolarization}) give
\begin{equation}
 \frac{\partial_tP_y^L}{\kb Q\delta T}
 =\av{\sum_n w_n(\partial_xE_n)d^\ell_{n,y}
       \cos\!\left[\frac{Qa\,\partial_xE_n\,t}{\hbar}\right]}.
 \label{supp:eq:packetreadout}
\end{equation}
Thus the short-transit-time polarization rate reproduces $\alpha^L_{yx}/\kb$ in Eq.~(\ref{supp:eq:general}), after integration by parts. Interchanging $x,y$ gives the other component. Their half difference yields
\begin{align}
 B_A^{\rm packet}(t)=\frac12\av{\sum_n w_n\left[
 (\partial_xE_n)d^\ell_{n,y}\cos\!\left(\frac{Qa\,\partial_xE_n t}{\hbar}\right)
 -(\partial_yE_n)d^\ell_{n,x}\cos\!\left(\frac{Qa\,\partial_yE_n t}{\hbar}\right)\right]},\\
 \lim_{Qv_{\max}t\to0}B_A^{\rm packet}(t)=\BA.
 \label{supp:eq:packetlimit}
\end{align}
The equality uses the full intrinsic dipole coefficient of the real-band model; it is not a conventional-current substitution. Physically, free streaming sorts packets with different angular dipoles and velocities, changing the local dipole polarization. This supplies a time-domain meaning for the dispersion--dipole correlation in the main text.

For $g=1$ meV, $\delta=2.5$ meV, and $T_0=300$ K, the $160^2$ calculation gives an interband time $0.620$ ps. At $Qa=10^{-3}$, the shortest grating transit time is $387$ ps. Between $3$ and $30$ ps, $B_A^{\rm packet}$ differs from $\BA=-3.51239\times10^{-5}$ by less than $0.17\%$ [main-text Fig.~1(c)]. Observing this interval requires a relaxation time longer than the observation time; no material linewidth is assumed in the calculation. The early-time coefficient equals the independently evaluated entropy-weighted kernel to $7\times10^{-20}$ at $Qa\,v_{\max}t/a=0$.

This protocol measures the first moment of angular density. A divergence-free transverse polarization need not produce a net bulk density signal; its readout resolves the angular distribution of the packets. Edge accumulation and mechanical torque additionally require the boundary, bath, and probe dynamics.

\subsection{Microscopic unitary readout of the resolved ensemble}
\label{supp:sec:microreadout}
The kinetic construction above can be tested without inserting the band-dipole expression into the readout. For each band $n$ and central momentum $\bm q$, prepare a normalized Gaussian packet of momentum standard deviation $\sigma_q$ and evolve its full orbital amplitudes with the original Hamiltonian:
\begin{equation}
 W_{n\bm q}(R,t)=\mathcal N\sum_{\bm p}
 e^{-|\bm p-\bm q|^2/(4\sigma_q^2)}e^{\ii\bm p\cdot R/a}
 e^{-\ii h(\bm p)t/\hbar}u_n(\bm p).
 \label{supp:eq:microevolution}
\end{equation}
The momentum sum is locally centered on $\bm q$ with the corresponding Bloch twist on the finite lattice; phases are aligned to $u_n(\bm q)$ within the Gaussian support. No interband dephasing assumption is needed for this band-prepared ensemble. At every time we measure
\begin{align}
 p_R&=W_R^\dagger W_R,\qquad \rho_R^\ell=W_R^\dagger\ell W_R,\nonumber\\
 X_i&=\sum_R(R_i/a)p_R,\qquad
 \mu_i=\sum_R(R_i/a-X_i)\rho_R^\ell.
 \label{supp:eq:micromoments}
\end{align}
Here $X_i$ is the dimensionless probability center and $\hbar a\mu_i$ is the measured internal angular first moment. Their time derivatives follow from $\dot W=-\ii hW/\hbar$ in the original orbital basis. Neither a velocity vertex nor an eigenvector derivative enters these measurements.

An incoherent grating of initial packet-center weights gives the Fourier component of the center-resolved polarization
\begin{equation}
 \frac{P_i^L(Q_j,t)}{\delta T}
 =\frac{\hbar}{a}\av{\sum_n\partial_Tn_n\,\mu_i(t)
       e^{-\ii QaX_j(t)}}.
 \label{supp:eq:micropolarization}
\end{equation}
We choose the initial centers to be zero before translating each packet into the ensemble. The complex Fourier amplitude is normalized so that its real part gives a cosine grating of amplitude $\delta T$. Therefore the measured tensor is
\begin{align}
 A_{ij}^{\rm micro}(Q,t)
 &=\operatorname{Re}\frac{\partial_tP_i^L(Q_j,t)}{-\ii Q\kb\delta T}\nonumber\\
 &=\frac{\hbar}{\kb}\operatorname{Re}\av{\sum_n\partial_Tn_n
 \left[\dot X_j\mu_i+\frac{\ii\dot\mu_i}{Qa}\right]
 e^{-\ii QaX_j}},\nonumber\\
 B_A^{\rm micro}&=(A_{yx}^{\rm micro}-A_{xy}^{\rm micro})/2.
 \label{supp:eq:microtensor}
\end{align}
This is a direct real-space measurement prescription. In the narrow-packet limit, $\mu_i\to d^\ell_{n,i}$ and $a\dot X_i\to v_{n,i}$, recovering Eq.~(\ref{supp:eq:packetlimit}); these relations are used only to compare the independent calculation with band kinetics.

At $T_0=300$ K, $g=1$ meV, and $\delta=2.5$ meV, a $32\times32$ central-momentum quadrature gives the initial values
\begin{center}
\begin{tabular}{cc}
\toprule
$\sigma_q$ & $B_A^{\rm micro}(0)$\\\midrule
0.12 & $-3.54438\times10^{-5}$\\
0.08 & $-3.52806\times10^{-5}$\\
0.04 & $-3.51654\times10^{-5}$\\\bottomrule
\end{tabular}
\end{center}
The narrowest packet differs by $0.12\%$ from the independently evaluated bulk coefficient, $-3.51239\times10^{-5}$. A linear extrapolation in $\sigma_q^2$ using the two smallest widths gives $-3.51270\times10^{-5}$, within $0.009\%$; this extrapolation is not used to plot the unitary data. Refining the central-momentum quadrature from $32^2$ to $48^2$ changes the narrow-packet initial value by less than $5\times10^{-8}$ relative. With $g=0$ and $\sigma_q=0.04$, the measured initial antisymmetric response is $3.62\times10^{-11}$ while both off-diagonal components are approximately $-2.560\times10^{-5}$. The finite-width and finite-$Qt$ readouts need not obey an exact zero away from the limiting coefficient.

Main-text Fig.~1(c) uses raw unitary results at $3$, $10$, and $30$ ps for each grating wave number. The first two times use $256^2$ cells; the $30$ ps point uses $384^2$ to avoid periodic wraparound of the translating packet. The $256^2$ result at $30$ ps is visibly affected by periodic wraparound. No fitted amplitude is applied. At $Qa=10^{-3}$ the measured $30$ ps value is $-3.51079\times10^{-5}$. The three sampled times lie within $0.12\%$ of the bulk coefficient. On the larger lattice, the probability within four cells of a boundary is below $2\times10^{-18}$ for every packet, and $|\dot\mu_i|<1.7\times10^{-15}$ ps$^{-1}$. The response arises from translating angular dipoles, rather than growth of an individual packet's moment.

The conditioning on packet center in Eq.~(\ref{supp:eq:micropolarization}) is essential. A single freely translating zero-monopole packet has a constant global angular first moment; the local polarization changes as packets enter and leave a resolved observation region. Summing their angular densities before extracting a moment loses that decomposition. In particular, a transverse polarization depending only on the drive coordinate is divergence free. An exploratory calculation of the total angular-density moment in an open strip, starting from a local-Gibbs covariance, displays coherent oscillations and does not reproduce this plateau. That calculation has a different preparation and readout. The unitary calculation establishes the resolved-ensemble response; it does not establish a reservoir-driven dc coefficient or a boundary torque.

\subsection{Field-correlation meaning of the cavity observable}
The vacuum window selects the optical field inside an actual cavity and fixes the rotation origin at its center. Restoring dimensional vacuum normalization, take the two-dimensional TM vector potential as $A_z$, its conjugate momentum as $\Pi_z=\epsilon_0\dot A_z$, and write $D_\varphi=\rho_x\nabla_{\rho,y}-\rho_y\nabla_{\rho,x}$. For a harmonic field at optical frequency $\omega$, $A_z=\operatorname{Re}[\mathcal A_z e^{-\ii\omega t}]$, $\mathcal E_z=\ii\omega\mathcal A_z$, the cycle-averaged canonical angular density is
\begin{equation}
 \overline{\mathcal L_z^{\rm cav}}
 =-\overline{\Pi_zD_\varphi A_z}
 =\frac{\epsilon_0\omega}{2}\operatorname{Im}(\mathcal A_z^*D_\varphi\mathcal A_z)
 =\frac{\epsilon_0}{2\omega}\operatorname{Im}(\mathcal E_z^*D_\varphi\mathcal E_z).
 \label{supp:eq:fieldreadout}
\end{equation}
Equation~(\ref{supp:eq:fieldreadout}) uses the vacuum volume density; integrating over the chosen mode-normalization length in the invariant $z$ direction gives the two-dimensional cell observable. For a stationary thermal field the products become cross-spectral field correlations and are integrated over frequency with the chosen spectral convention. For a broadband transient, the equal-time symmetrized $\Pi_z,A_z$ correlation retains the interfrequency terms. Thus phase-sensitive field-correlation tomography provides an in-principle readout of the specified cavity density and its spatial first moment. Orbital-resolved polarimetry has already reconstructed amplitudes and phases in a multimode photonic lattice \cite{Guillot2025}; adapting such reconstruction to the present thermal protocol is an experimental implementation question.

The chosen density follows from the scalar canonical action and generates spatial rotations, which need not be a symmetry of an elliptical cavity. Angular exchange with that cavity appears through its torque. The distinction between canonical optical momentum and the total mechanical balance in dispersive matter \cite{Bliokh2017} is addressed here by specifying a vacuum-field observable. Enlarging the window into the host changes that observable. Mirror-preserving window changes retain the parity-channel selection rule, but its magnitude and sign depend on the angular matrix elements. No window-independence assumption enters the result.

\subsection{Isolation scale of the lowest Maxwell band}
The reference field spectrum gives $\min_{\bm q}[E_1(\bm q)-E_0(\bm q)]=4.67303$ meV on its $64\times64$ shifted momentum grid and $40\times40$ spatial mesh. The sampled minimum lies next to the zone corner, at $q_x=q_y=-63\pi/64$, with energies $15.87468$ and $20.54771$ meV. The $48\times48$ spatial calculation, sampled on a $32\times32$ momentum grid, gives $4.67880$ meV. These are sampled minima, not certified bounds on the continuous Brillouin zone; the two calculations also use different momentum meshes. The approximately $4.67$ meV separation corresponds to an interband time of $0.14$ ps for the lowest band. It supports the isolated-band low-temperature treatment without specifying a material relaxation rate or the gaps between higher modes.

\subsection{Thermal activation and signal scale}
For an isolated quadratic minimum with $C_0\ne0$, Eq.~(\ref{supp:eq:laplace}) supplies a nonzero leading intrinsic response as $T\to0$. Higher bands contribute virtually to $C_0$ while their thermal populations are exponentially smaller. The asymptotic prefactor and the finite-temperature magnitude must be distinguished: extrapolating $C_0e^{-E_{\min}/(\kb T)}$ to $10$ K gives $6.09\times10^{-10}$, about $26.3$ times the direct integral. The quoted $2.32\times10^{-11}$ value and main-text temperature curve come from the full Brillouin-zone quadrature, not this extrapolation.

For dimensional orientation, a gradient of $1$ K/mm gives the intrinsic antisymmetric current normalization $J_A^L=\kb\BA|\nabla T|$. The minimal model at $300$ K yields $|J_A^L|=4.85\times10^{-25}$ J/m, while the field result at $10$ K yields $3.20\times10^{-31}$ J/m. Multiplying by a $1$ mm width gives angular rates of $4.60\times10^6\hbar/{\rm s}$ and $3.03\hbar/{\rm s}$, respectively. These are conversions of the intrinsic coefficient into angular-current units; they do not specify a detector torque or its noise floor.

The distinction is also explicit under a uniform positive energy rescaling. At fixed dimensionless Hamiltonian and angular observable, let $E_n\mapsto sE_n$ with $s>0$. The eigenvectors, dipoles, and kernels are unchanged, so
\begin{equation}
 B_A(T;sE_0)=B_A(T/s;E_0),\qquad
 C_0(sE_0)=\frac{sE_{\min}\mathcal F_0(\Gamma)}{2\pi\sqrt{\det(sM)}}=C_0(E_0).
 \label{supp:eq:rescale}
\end{equation}
The second equality uses two spatial dimensions. Thus absolute temperature depends on the energy scale chosen for an ideal model, while the dimensionless geometric response and activation prefactor do not. Uniform rescaling shifts a response curve along the temperature axis; it does not increase its magnitude at fixed $\kb T/E_0$ or improve isolation from other bands. In the Drude field model, keeping the dimensionless geometry and $\kappa=\omega_pR_x/c$ fixed entails correlated changes of length and plasma frequency. Equation~(\ref{supp:eq:rescale}) does not assert that an arbitrary rescaling is available in a particular material.

The field result concerns the specified scalar TM polarization and the collisionless current. Extending it to higher reduced temperatures requires consistent treatment of crossing bands; a stationary detector signal additionally requires relaxation and probe vertices. TE modes have different Maxwell and normalization operators, while a three-dimensional slab also has radiation channels. The geometric selection rule applies wherever its real-band, locality, and isolation assumptions hold; the numerical coefficient must be recomputed for each such implementation. This separates the established mechanism from the device-design extensions.


\begin{thebibliography}{99}
\bibitem{Onoda2004} M. Onoda, S. Murakami, and N. Nagaosa, Hall effect of light, \href{https://doi.org/10.1103/PhysRevLett.93.083901}{Phys. Rev. Lett. \textbf{93}, 083901 (2004)}.
\bibitem{Bliokh2015} K. Y. Bliokh, F. J. Rodr\'{i}guez-Fortu\~no, F. Nori, and A. V. Zayats, Spin--orbit interactions of light, \href{https://doi.org/10.1038/nphoton.2015.201}{Nat. Photonics \textbf{9}, 796--808 (2015)}.
\bibitem{Athira2021} Athira B. S., S. Mukherjee, A. Laha, K. Bar, D. Nandy, and N. Ghosh, Experimental observation of the orbital Hall effect of light through pure orbit--orbit interaction for randomly and radially polarized vortex beams, \href{https://doi.org/10.1364/JOSAB.426304}{J. Opt. Soc. Am. B \textbf{38}, 2180--2186 (2021)}.
\bibitem{Bernevig2005} B. A. Bernevig, T. L. Hughes, and S.-C. Zhang, Orbitronics: The intrinsic orbital current in $p$-doped silicon, \href{https://doi.org/10.1103/PhysRevLett.95.066601}{Phys. Rev. Lett. \textbf{95}, 066601 (2005)}.
\bibitem{Go2018} D. Go, D. Jo, C. Kim, and H.-W. Lee, Intrinsic spin and orbital Hall effects from orbital texture, \href{https://doi.org/10.1103/PhysRevLett.121.086602}{Phys. Rev. Lett. \textbf{121}, 086602 (2018)}.
\bibitem{Salemi2022} L. Salemi and P. M. Oppeneer, First-principles theory of intrinsic spin and orbital Hall and Nernst effects in metallic monoatomic crystals, \href{https://doi.org/10.1103/PhysRevMaterials.6.095001}{Phys. Rev. Materials \textbf{6}, 095001 (2022)}.
\bibitem{Li2026} W.-Y. Li, C.-C. Chang, S.-Y. Huang, and D. Qu, Transport evidence for the orbital Nernst effect, \href{https://doi.org/10.1038/s41467-026-77878-2}{Nat. Commun. (2026), doi:10.1038/s41467-026-77878-2}.
\bibitem{Masuda2026} Y. Masuda \emph{et al.}, Observation of the orbital Nernst effect, \href{https://arxiv.org/abs/2609.27604}{arXiv:2609.27604 (2026)}.
\bibitem{Cheng2016} R. Cheng, S. Okamoto, and D. Xiao, Spin Nernst effect of magnons in collinear antiferromagnets, \href{https://doi.org/10.1103/PhysRevLett.117.217202}{Phys. Rev. Lett. \textbf{117}, 217202 (2016)}.
\bibitem{Zyuzin2016} V. A. Zyuzin and A. A. Kovalev, Magnon spin Nernst effect in antiferromagnets, \href{https://doi.org/10.1103/PhysRevLett.117.217203}{Phys. Rev. Lett. \textbf{117}, 217203 (2016)}.
\bibitem{Weissenhofer2026} M. Wei\ss enhofer, M. S. Mrudul, S. Mankovsky, and P. M. Oppeneer, Magnon orbital Nernst effect in altermagnets, \href{https://doi.org/10.1038/s41535-026-00853-z}{npj Quantum Mater. \textbf{11}, 25 (2026)}.
\bibitem{Park2020} S. Park and B.-J. Yang, Phonon angular momentum Hall effect, \href{https://doi.org/10.1021/acs.nanolett.0c03220}{Nano Lett. \textbf{20}, 7694--7699 (2020)}.
\bibitem{Lopez2026} D. A. Bustamante Lopez, V. Brehm, and D. M. Juraschek, Atomistic theory of the phonon angular momentum Hall effect, \href{https://arxiv.org/abs/2604.01899}{arXiv:2604.01899 (2026)}.
\bibitem{Ben2016} P. Ben-Abdallah, Photon thermal Hall effect, \href{https://doi.org/10.1103/PhysRevLett.116.084301}{Phys. Rev. Lett. \textbf{116}, 084301 (2016)}.
\bibitem{Chen2022} P. Y. Chen, C. Khandekar, R. Ayash, Z. Jacob, and Y. Sivan, Thermal emission of spinning photons from temperature gradients, \href{https://doi.org/10.1103/PhysRevApplied.18.014052}{Phys. Rev. Applied \textbf{18}, 014052 (2022)}.
\bibitem{Ben2025} P. Ben-Abdallah, Inverse spin thermal Hall effect in nonreciprocal photonic systems, \href{https://doi.org/10.1103/PhysRevLett.134.113804}{Phys. Rev. Lett. \textbf{134}, 113804 (2025)}.
\bibitem{Ben2026} P. Ben-Abdallah, Radiative spin caloritronics, \href{https://arxiv.org/abs/2607.22845}{arXiv:2607.22845 (2026)}.
\bibitem{Dehaghi2026} A. K. Dehaghi and L. Zhu, Near-field photon Nernst effect, \href{https://doi.org/10.1103/z52f-tjqt}{Phys. Rev. Lett. \textbf{137}, 116902 (2026)}.
\bibitem{Milicevic2017} M. Mili\'cevi\'c \emph{et al.}, Orbital edge states in a photonic honeycomb lattice, \href{https://doi.org/10.1103/PhysRevLett.118.107403}{Phys. Rev. Lett. \textbf{118}, 107403 (2017)}.
\bibitem{Shi2006} J. Shi, P. Zhang, D. Xiao, and Q. Niu, Proper definition of spin current in spin-orbit coupled systems, \href{https://doi.org/10.1103/PhysRevLett.96.076604}{Phys. Rev. Lett. \textbf{96}, 076604 (2006)}.
\bibitem{Xiao2021} C. Xiao and Q. Niu, Conserved current of nonconserved quantities, \href{https://doi.org/10.1103/PhysRevB.104.L241411}{Phys. Rev. B \textbf{104}, L241411 (2021)}.
\bibitem{Koyama2025} S. Koyama and J. Nasu, Formulation of spin Nernst effect for spin-nonconserving insulating magnets, \href{https://doi.org/10.1103/8xgc-d9ch}{Phys. Rev. B \textbf{112}, 014447 (2025)}.
\bibitem{Culcer2004} D. Culcer, J. Sinova, N. A. Sinitsyn, T. Jungwirth, A. H. MacDonald, and Q. Niu, Semiclassical spin transport in spin-orbit-coupled bands, \href{https://doi.org/10.1103/PhysRevLett.93.046602}{Phys. Rev. Lett. \textbf{93}, 046602 (2004)}.
\bibitem{Luttinger1964} J. M. Luttinger, Theory of thermal transport coefficients, \href{https://doi.org/10.1103/PhysRev.135.A1505}{Phys. Rev. \textbf{135}, A1505 (1964)}.
\bibitem{Qin2011} T. Qin, Q. Niu, and J. Shi, Energy magnetization and the thermal Hall effect, \href{https://doi.org/10.1103/PhysRevLett.107.236601}{Phys. Rev. Lett. \textbf{107}, 236601 (2011)}.
\bibitem{Matsumoto2011} R. Matsumoto and S. Murakami, Rotational motion of magnons and the thermal Hall effect, \href{https://doi.org/10.1103/PhysRevB.84.184406}{Phys. Rev. B \textbf{84}, 184406 (2011)}.
\bibitem{SM} See the Supplemental Material appended below for current conventions, selection-rule derivations, canonical-field reduction, numerical convergence, projection identities, and microscopic readout details.
\bibitem{Provost1980} J. P. Provost and G. Vall\'{e}e, Riemannian structure on manifolds of quantum states, \href{https://doi.org/10.1007/BF02193559}{Commun. Math. Phys. \textbf{76}, 289--301 (1980)}.
\bibitem{Ahn2019} J. Ahn, S. Park, and B.-J. Yang, Failure of Nielsen--Ninomiya theorem and fragile topology in two-dimensional systems with space-time inversion symmetry: Application to twisted bilayer graphene at magic angle, \href{https://doi.org/10.1103/PhysRevX.9.021013}{Phys. Rev. X \textbf{9}, 021013 (2019)}.
\bibitem{Bouhon2020} A. Bouhon, T. Bzdu\v{s}ek, and R.-J. Slager, Geometric approach to fragile topology beyond symmetry indicators, \href{https://doi.org/10.1103/PhysRevB.102.115135}{Phys. Rev. B \textbf{102}, 115135 (2020)}.
\bibitem{Bliokh2010} K. Y. Bliokh, M. A. Alonso, E. A. Ostrovskaya, and A. Aiello, Angular momenta and spin-orbit interaction of nonparaxial light in free space, \href{https://doi.org/10.1103/PhysRevA.82.063825}{Phys. Rev. A \textbf{82}, 063825 (2010)}.
\bibitem{Bliokh2017} K. Y. Bliokh, A. Y. Bekshaev, and F. Nori, Optical momentum, spin, and angular momentum in dispersive media, \href{https://doi.org/10.1103/PhysRevLett.119.073901}{Phys. Rev. Lett. \textbf{119}, 073901 (2017)}.
\bibitem{Go2024} D. Go, H.-W. Lee, P. M. Oppeneer, S. Bl\"ugel, and Y. Mokrousov, First-principles calculation of orbital Hall effect by Wannier interpolation: Role of orbital dependence of the anomalous position, \href{https://doi.org/10.1103/PhysRevB.109.174435}{Phys. Rev. B \textbf{109}, 174435 (2024)}.
\bibitem{Guillot2025} M. Guillot \emph{et al.}, Measuring non-Abelian quantum geometry and topology in a multi-gap photonic lattice, \href{https://arxiv.org/abs/2511.03894}{arXiv:2511.03894 (2025), version 2 (2026)}.
\bibitem{Colpa1978} J. H. P. Colpa, Diagonalization of the quadratic boson Hamiltonian, \href{https://doi.org/10.1016/0378-4371(78)90160-7}{Physica A \textbf{93}, 327--353 (1978)}.
\bibitem{Zhang2022} Y.-M. Zhang, T. Zhu, Z.-Q. Zhang, and J.-S. Wang, Microscopic theory of photon-induced energy, momentum, and angular momentum transport in the nonequilibrium regime, \href{https://doi.org/10.1103/PhysRevB.105.205421}{Phys. Rev. B \textbf{105}, 205421 (2022)}.
\bibitem{Wang2023} J.-S. Wang, J. Peng, Z.-Q. Zhang, Y.-M. Zhang, and T. Zhu, Transport in electron-photon systems, \href{https://doi.org/10.1007/s11467-023-1260-z}{Front. Phys. \textbf{18}, 43602 (2023)}.
\end{thebibliography}
\end{document}